\documentclass[final,5p,times,authoryear]{elsarticle}

\usepackage{epsfig}
\usepackage{graphicx}

\newcommand{\ba}{\begin{eqnarray}}
\newcommand{\ea}{\end{eqnarray}}
\newcommand{\beq}{\begin{equation}}
\newcommand{\eeq}{\end{equation}}

\newcommand{\apj}{ Astrophys.\ J.}
\newcommand{\apjl}{ Astrophys.\ J.\ Letters}
\newcommand{\aap}{ Astron.\ Astrophys.}
\newcommand{\prd}{ Phys.\ Rev.\ D}

\def\e{\epsilon}

\def\ep{\epsilon^\prime}

\newcommand{\arrowsim}{\lower.5ex\hbox{$\; \buildrel \longrightarrow \over {\small{b = 1}} \;$}}

\newcommand{\arrsim}{\lower.5ex\hbox{$\; \buildrel \longrightarrow \over {\small{f_0 = 1/3}} \;$}}

\usepackage{amssymb}

\journal{Journal of High Energy Astrophysics}

\begin{document}

\begin{frontmatter}



\title{Photopion Production in Black-Hole Jets \\and Flat-Spectrum Radio Quasars as PeV Neutrino Sources}


\author{Charles D. Dermer,$^1$ Kohta Murase,$^2$ \& Yoshiyuki Inoue$^3$}

\address{$^1$Space Science Division, 
Code 7653, Naval Research Laboratory, Washington, DC 20375-5352; charles.dermer@nrl.navy.mil\\
$^2$ Insitute for Advanced Study, 1 Einstein Dr., Princeton, NJ 08540, USA; murase@ias.edu\\
$^3$ Institute of Space and Astronautical Science, Japan AeroSpace Exploration Agency, 3-1-1 Yoshinodai, Chuoku, Sagamihara, Kanagawa, 252-5210, Japan}

\begin{abstract}
The IceCube collaboration has reported { neutrinos with energies between $\sim 30$ TeV and a few PeV that are significantly enhanced over the cosmic-ray induced atmospheric background.} Viable high-energy neutrino sources  must contain very high-energy and ultra-high energy cosmic rays while efficiently making PeV neutrinos. Gamma-ray Bursts (GRBs) and blazars have been considered as candidate cosmic-ray accelerators. GRBs, including low-luminosity GRBs, can be efficient PeV
neutrino emitters for low bulk Lorentz factor outflows, although the
photopion production efficiency needs to be tuned to simultaneously
explain ultra-high-energy cosmic rays.
Photopion production efficiency of cosmic-rays accelerated in the inner jets of flat spectrum radio quasars (FSRQs)  is $\sim 1$ -- 10\%  due to interactions with photons of the broad-line region (BLR), whereas BL Lac objects are not effective PeV neutrino sources due to the lack of external radiation fields.  Photopion threshold effects with BLR photons suppress neutrino production below $\sim 1$ PeV, which imply that neutrinos from other sources would dominate over the diffuse neutrino intensity at sub-PeV energies. Reduction of the $\gg$ PeV neutrino flux can be expected when curving cosmic-ray proton distributions are employed. Considering a log-parabolic function to describe the cosmic-ray  distribution, we discuss possible implications for particle acceleration in black-hole jets.  Our results encourage a search for IceCube PeV neutrino events  associated with $\gamma$-ray loud FSRQs using Fermi-LAT data. 
In our model, as found in our previous work, the neutrino flux is suppressed below 1 PeV, which can be tested with increased IceCube exposure.
\end{abstract}


\end{frontmatter}


\section{Introduction}

The IceCube Collaboration has reported evidence for
extragalactic neutrinos \citep{IceCube2013}, which opens up an important multi-messenger connection
between photons, neutrinos, and high-energy cosmic rays. 
Using two years of IceCube data in its 79 and 86 string configuation, 
28 fully contained events were identified between
30 and 1200 TeV, of which 21 were shower-like, and the remainder track-like. 
Using three years of data, 37 events are reported, with 
28 shower-like and 9 track-like events  \cite{aar14}. 
All of the highest energy neutrinos, 
with energies of 1040, 1140 TeV, and 2004 TeV and energy uncertainties of $\approx 12$\%, are shower-like. The ratio of showers and tracks is a consequence of the larger effective area of IceCube for $\nu_e$ interactions \citep{IceCubeprl2013}, different neutrino-flavor opacities through the Earth, and the analysis requirement that the events are fully contained.

The combined significance of the data is  $\approx 5.7\sigma$  over
predicted background, systematic uncertainties and uncertain charm contribution, with the 
significance of separate low-energy (LE, $\approx 25$ -- 500 TeV) and high-energy (HE; $\gtrsim$ 0.5 PeV)
neutrino enhancements less significant. Indeed, the available evidence 
for a suppression of neutrino production near $\sim 0.5$ PeV is not statistically significant,
either in the two-year or three-year data sets.
At higher energies, between $\approx 2$ and 10 PeV,
3 -- 6 neutrinos were predicted for a proton spectrum with index $=-2$, whereas none were reported  \citep{IceCube2013,aar14}. 
Thus the existence of a high-energy
cutoff above the energies of the two $\sim 1$ PeV events and the 
recently announced $\sim 2$ PeV event \citep{aar14,klein13,ancho13}, is statistically 
favored though not definitely established.  
The  neutrino flux
is adequately fit with a $-2$ neutrino spectrum down to low energies \citep{IceCube2013}, 
and is inconsistent with a diffuse cosmogenic origin of neutrinos from UHECRs in the intergalactic medium \cite{aar13,rou13,lah13}.  

Here we consider whether neutrino production with photons of the broad-line region (BLR) of flat spectrum radio quasars (FSRQs),  can account for the features of the PeV neutrinos detected by IceCube. This process was first considered in detail by \citet{ad01}, though suggestions of neutrino production from FSRQs were made earlier \cite{msb92,rm98}.
If atomic-line radiation in the BLR dominates neutrino production through photopion processes, suppression of the neutrino flux from FSRQs at energies $\lesssim 1$ PeV is expected.
These cutoffs are easily understood by noting that for threshold pion production, $\gamma_p\epsilon_* \gtrsim m_\pi/m_e \cong 300$.
For neutrinos formed with $\approx 5$\% of the incident proton energy, then a cutoff is expected
 at $E_\nu \approx 0.05 m_pc^2 (m_\pi/ m_e \epsilon_*) \sim 1$ PeV, taking  $\epsilon_*\approx 2\times 10^{-5}$ for the Lyman $\alpha$ photon energy. 

In this paper, we study the emission of HE neutrinos produced by photopion processes in extragalactic black-hole jet sources, focusing in particular on FSRQs. Inefficient Fermi acceleration competing with strong photohadronic energy losses due to atomic-line photons in the BLR of FSRQs is shown to give proton distributions with cutoffs at $\approx 10^{16}$ eV. In related work \cite{mid14}, we calculate the diffuse neutrino background from the superposition of distant blazars, where we also find a suppression of the neutrino flux below $1$ PeV.

\section{Photopion efficiency with internal target electron synchrotron photons}

Photopion production of high-energy ($\approx 10^{14}$ -- $10^{17}$ eV) cosmic rays in the intense BLR and internal radiation fields of blazars is more energetically efficient than secondary nuclear production in proton-ion collisions, provided the threshold for pion production is achieved  \citep{ad03}.  To calculate the proton energy-loss timescale through photopion losses, we use the approximation $K_{\phi\pi}(\bar\epsilon_r) \sigma_{\phi\pi}(\bar\epsilon_r ) \cong \hat\sigma H(\bar\epsilon_r-\bar\epsilon_{thr})$ for the product of the inelasticity and photopion production cross section, where $\bar\epsilon_r$ is the invariant 
photon energy in the particle rest frame. Here $\hat\sigma = 70 \,\mu$b, and $\bar\epsilon_{thr}\cong 400$.  The Heaviside function $H(x) = 1$ for $x\geq 0$ and $H(x) = 0$ otherwise.

The timescale $t^\prime_{\phi\pi}(\gamma_p^\prime)$ for a proton of energy $m_pc^2 \gamma_p^\prime$ to lose energy through photopion production is given by  
\begin{equation}
t^{\prime -1}_{\phi\pi}(\gamma_p^\prime) = c\hat\sigma \int_{\bar\epsilon_{thr}/2\gamma_p^\prime }^\infty d\epsilon^\prime \,n_{ph}^\prime (\ep ) \,[1-({\bar\epsilon_{thr}\over 2\gamma_p^\prime\ep})^2]\;
\label{tprime-1}
\end{equation}
\citep{ste69}, 
with primes referring to comoving fluid-frame quantities. The term $n_{ph}^\prime(\ep )$ is the comoving spectral photon number density, $\ep$ is 
the comoving dimensionless photon energy, and the pitch-angle diffusion timescale of the particles is assumed to be rapid enough to isotropize the particle distribution in the fluid frame.

We adopt expressions for the nonthermal 
synchrotron luminosity radiated by
an isotropic comoving electron distribution $\gamma^{\prime 2}_e N^\prime_e(\gamma^{\prime}_e)$ described by a log-parabola function, 
where $\gamma^\prime_e$ is the electron Lorentz factor in the comoving frame \citep{der14}. 
In this approximation, the synchrotron luminosity spectrum
\begin{equation}
\epsilon L_{syn}(\epsilon )= \upsilon x^{1-\hat b {\rm ln}x}\;,
\label{eLsyne}
\end{equation} 
where $\upsilon =10^{-1/4b} (\nu L_\nu^{pk,syn})$,
$b = \hat b \ln 10$ is the log-parabola width parameter of the electron distribution, 
$x = \sqrt{\epsilon/\epsilon_{pk} } = \sqrt{\ep/\ep_{pk}}$,  $\epsilon^\prime = \epsilon/\delta_{\rm D}$,  $\epsilon^\prime_{pk} = \epsilon_{pk}/\delta_{\rm D}$, and $\delta_{\rm D}$ is the Doppler factor.  The peak synchrotron luminosity $\nu L_\nu^{pk,syn}$ 
at peak synchrotron frequency $\epsilon_s = 10^{1/b}\epsilon_{pk} = 8.1\times 10^{-7}\nu_{pk,14}$ 
 is derived directly from the data for a source at redshift $z$. In the blob formulation, 
\begin{equation}
n^\prime_{ph}(\ep ) = {\ep u^\prime(\ep )\over m_ec^2 \epsilon^{\prime 2}} = {\epsilon L_{syn}(\epsilon )\over 4\pi m_ec^3 \epsilon^{\prime 2} r_b^{\prime 2} \delta_{\rm D}^4 f_0 }\;,
\label{nprimephep}
\end{equation}
where  $f_0\approx 1/3$. In the blast-wave formulation, $f_0 \cong 1$, $\Gamma\cong \delta_{\rm D}$, and $n^\prime_{ph}(\ep ) = {\epsilon L_{syn}(\epsilon )/ 4\pi m_ec^3 \epsilon^{\prime 2} r^{2} \Gamma^2 }$, with $r\approx c \Gamma^2 t_{var}$, leading to effectively equivalent results \cite{dm09,mid14}. For blazar calculations using the blob formulation, $\delta_{\rm D} \cong \Gamma$ is assumed. { Synchrotron self-absorption is not important for PeV neutrino production in blazars and GRBs (see Appendix A), and internal and source $\gamma\gamma$ opacity is less important for the neutrino spectrum than the $\gamma$-ray spectral energy distribution
(SED) \citep{der12,der14}. }

The photopion radiative efficiency of ultra-high-energy 
cosmic ray (UHECR; energies $\gtrsim 10^{17}$ eV) 
protons with Lorentz factor $\gamma_p\cong \delta_{\rm D}\gamma_p^\prime$
is defined by the expression 
$\eta_{\phi\pi} \equiv t^\prime_{dyn}/t^\prime_{\phi\pi}(\gamma_p)$, 
where $t_{dyn}^\prime \cong \delta_{\rm D} t_{var} \cong \Gamma t_{var}$ is the comoving dynamical timescale.
Eqs.\ (\ref{tprime-1}) -- (\ref{nprimephep}) imply that the efficiency of UHECR protons to lose
energy through photopion production with internal synchrotron photons is
\begin{equation}
\eta^{int}_{\phi\pi} = \eta_s \, I_s(\bar x)\;,
\label{etaint}
\end{equation} 
where
\begin{equation}
I_s(\bar x) \equiv \int_{\bar x}^\infty dx\,x^{-2-\hat b \ln x} (1 - {\bar x^4\over x^4})\;
\,,
\label{Isbarx}
\end{equation}
\begin{equation}
\bar x \equiv \delta_{\rm D}\sqrt{\bar\epsilon_{thr}/ 2\gamma_{p} \epsilon_{pk}},
\label{barx}
\end{equation}
\begin{equation}
\eta_s = {\hat \sigma 10^{3/4b}(\nu L_\nu^{pk,syn})\over 2\pi m_ec^4 t_{var} \delta^4_{\rm D} f_0 \epsilon_{s} }
\cong  1.5\times 10^4 \,{10^{3/4b} L_{48}\over t_{var}({\rm s}) \delta^4_{\rm D}f_0 \epsilon_{s}}\;,
\label{etas}
\end{equation}
and $L_{48} \equiv \nu L_\nu^{pk,syn}/(10^{48}{\rm~erg~s}^{-1})$. This expression likewise applies
to a spherical blast-wave geometry, letting $\delta_{\rm D}\rightarrow \Gamma$ and taking $f_0\approx 1$.
Note that $I_s(\bar x)\rightarrow 10^{1/4b}\sqrt{{\pi\ln 10/ b}}$ in the limit $\bar x \ll 1$ \citep{der14}, 
and $I_s(\bar x)\approx 4\bar x^{1-k}/[(k-1)(k+3)]$ in the limit $\bar x \gg 1$, where $k \equiv 2+\hat b \ln \bar x$.
Fig.\ 1 shows a numerical integration of $I_s(\bar x )$, eq.\ (\ref{Isbarx}), for different values of $b$, compared
to the $\bar x\gg 1$ asymptotes. 
When $\bar x \ll 1$, corresponding to large $\gamma_p$, the production efficiency, eq.\ (\ref{etaint}), approaches
a constant value. 

\begin{figure}[t]
\label{Fig1}
\centering
\includegraphics[scale=0.5]{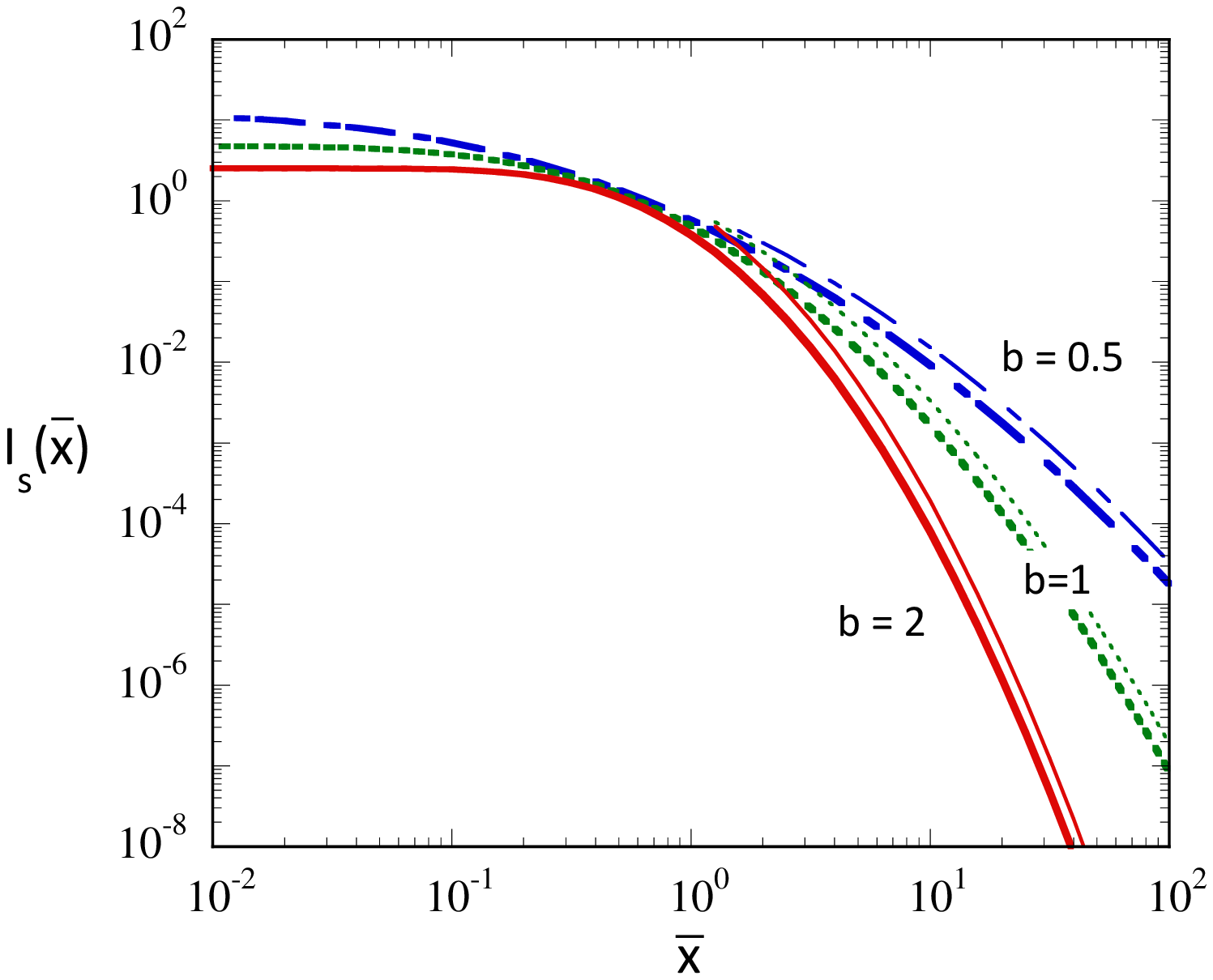}
\vskip0.1in 
\noindent{{\bf Fig.\ 1} Integral $I_s(\bar x )$, from eq.\ (\ref{Isbarx}). Thin curves show the high-energy approximation given in the text.}
\end{figure}

\section{Internal synchrotron photopion production efficiency of black-hole jet sources}

Fig.\ 2 shows calculations of photopion production efficiencies
using
eqs.\ (\ref{etaint}) -- (\ref{etas}) for different classes of black-hole jet sources, using
characteristic parameter values given in Table 1. Here we take $b = 1$. 
A value of $b$ near unity is implied for the electron distribution
by fitting the nonthermal synchrotron emission of 3C 279 \cite{der14}, and we assume 
it also applies for the proton distribution.  The Doppler factors are similar to values
implied by equipartition leptonic models \cite{der14}.\footnote{ By ``equipartition" we mean
equality of the energy densities of the magnetic-field and nonthermal electrons, that
is, $u^\prime_{B^\prime} = \zeta_e u_e^\prime$, with $\zeta_e \cong 1$. If instead the
 all-particle energy density $u^\prime_{par}= u_e^\prime + u^\prime_{p/nuc}$ is related
to the magnetic field according to the equipartition 
relation $u^\prime_{par} = \zeta_{eq}u^\prime_{B^\prime}$, where
$u^\prime_{p,nuc}$ is the energy density of hadrons in the blazar jet,
 this modifies the equipartition 
expressions by replacing $\zeta_e$ with $\zeta_{eq}/(1+f_{he}$) in the relation 
$u_e^\prime = \zeta_e u^\prime_{B^\prime}$, 
where the hadron-electron loading factor $f_{he} \equiv u^\prime_{p/nuc}/u_e^\prime$. For large hadronic loading,  
 when $f_{he} \gg 1$, the power requirements increase with accompanying spectral effects on the Compton component due to smaller values of $\delta_{\rm D}$ and larger values of $B^\prime$.} 
Only the synchrotron radiation field is assumed to be important for photopion production in 
the sources considered in Fig.\ 2; 
the higher-energy synchrotron self-Compton radiation fields have too few photons to be effective targets 
for photopion production. What is most notable is the 
extreme sensitivity of the efficiency to $\Gamma$ or $\delta_{\rm D}$, with $\eta^{int}_{\phi\pi} \propto \Gamma^{-4}$ at large proton energies. For LGRBs and SGRBs, low ($\Gamma\sim 100$)  outflows are potentially much more neutrino luminous than for high ($\Gamma \sim 1000$) bursts. { Fermi-LAT results suggest that the most powerful GRBs are those with the largest bulk Lorentz factor outflows \citep{cen11}, but to optimize neutrino luminosity, a smaller value of $\Gamma$ is required (Appendix B).}  This suggests examining neutrino
production from GRBs that can be shown to have small $\Gamma$ factors, e.g., GRB 090926A whose 
Fermi-LAT spectrum shows a cutoff that suggests that $\Gamma\sim 200$ -- 700 \cite{090926A}. 

The photomeson efficiency of LLGRBs is poorly known due to the large uncertainty in 
determining $\Gamma$ and $t_{var}$. For a hydrodynamic jet to penetrate the star, 
$\Gamma\sim 5$ is suggested \cite{tom07}. The synchrotron self-absorption interpretation 
of the low-energy spectrum also indicates that $\Gamma\sim 5$ and dissipation radii around the
photosphere \cite{ggt07}. Values of $\Gamma \sim 5$ -- 20 are considered in  \citep{mur06}; see 
also \cite{gz07,liu11}.
Related to the LLGRBs are shock-breakout GRBs, where the dissipation is caused by transrelativistic 
ejecta with $\Gamma\sim$ a few, and GRBs where neutrino production takes place in the star \cite{mi13,kas13}. 
We consider a broad range of $\Gamma$ between $\sim 2$ and 
30, and take $t_{var} = 100$ s.

\begin{table}[t]
\begin{center}
\noindent\centering{\bf Table 1.} {Parameters for Different Classes of Relativistic \\Black-Hole Jet Systems}
\begin{tabular}{cccccc}
\hline
$\# $ & Source & $\nu L_\nu^{pk,syn}$  & $t_{var}$  & $\delta_{\rm D} \cong \Gamma$   & $\nu_{pk,14}$  \\
 & Class  & ($10^{48}$ erg s$^{-1}$)  & (s)  &   & ($10^{14}$ Hz) \\
\hline
1a,b & LGRB$^a$ & 1000 & 0.1 & 100, 1000 & $2\times 10^5$\\
2a,b & SGRB$^b$ & 1000 & $10^{-3}$ & 100, 1000 & $10^6$\\
3a,b & LLGRBs$^c$ & 0.1 & 100 & 2, 30 & $10^4$\\ 
4a & BL Lac$^d$ & 0.001 & $10^5$ & 5 & $10^2$\\ 
4b & BL Lac$^d$ & 0.003 & 100 & 100 & $10^3$\\ 
5a & FSRQ$^e$ & 0.03 & $10^6$ & 10 & $0.1$\\ 
5b & FSRQ & 0.1 & $10^4$ & 30 & $0.1$\\ 
\hline\end{tabular}
\label{table1}
\end{center}
\noindent $^a$ Long Duration GRB\\
\noindent $^b$ Short Duration GRB\\
\noindent $^c$ Low-luminosity GRBs; \citep{mur06}\\
\noindent $^d$ High-synchrotron peaked BL Lac object\\
\noindent $^e$ Flat Spectrum Radio Quasars\\
\end{table}

\begin{figure}[t]
\centering
{\includegraphics[scale=0.5]{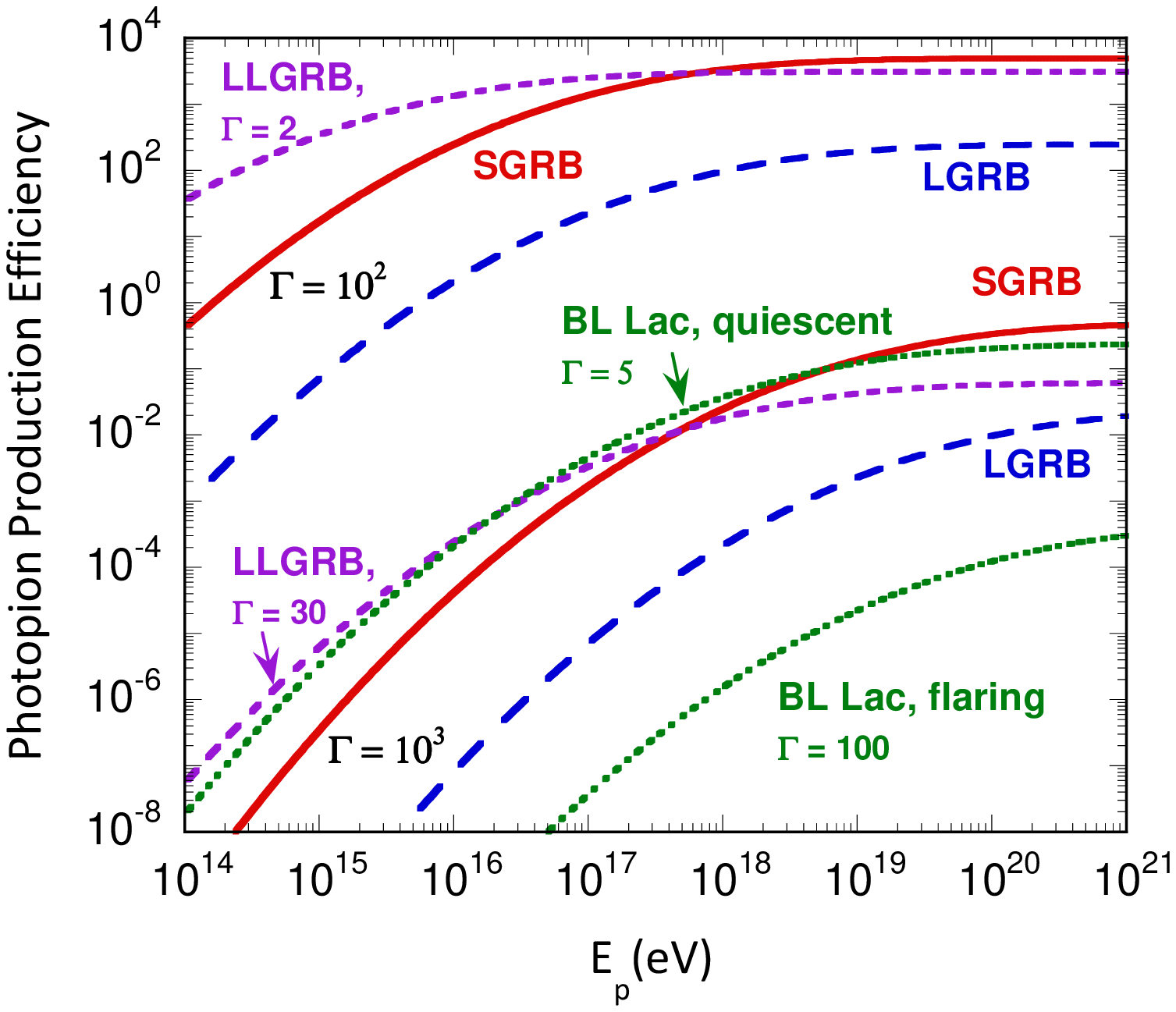}}
\vskip0.1in\flushleft
\noindent {\bf Fig. 2.} Photopion production efficiency { as a function of escaping proton energy $E_p = \Gamma m_p \gamma_p^\prime$,} presented in terms of the ratio of the dynamical and energy-loss timescales 
using parameters from Table 1 for long soft GRBs (LGRBs), short hard GRBs (SGRBs), 
low-luminosity GRBs (LLGRBs), and high-synchrotron peaked (HSP) BL Lac objects. For the efficiency calculations,
 $t_{dyn}^\prime = \Gamma t_{var} = \delta_{\rm D} t_{var}$ for photopion production with internal synchrotron photons.
 \label{Fig2}
\end{figure}


\begin{figure}[t]
\centering
{\includegraphics[scale=0.55]{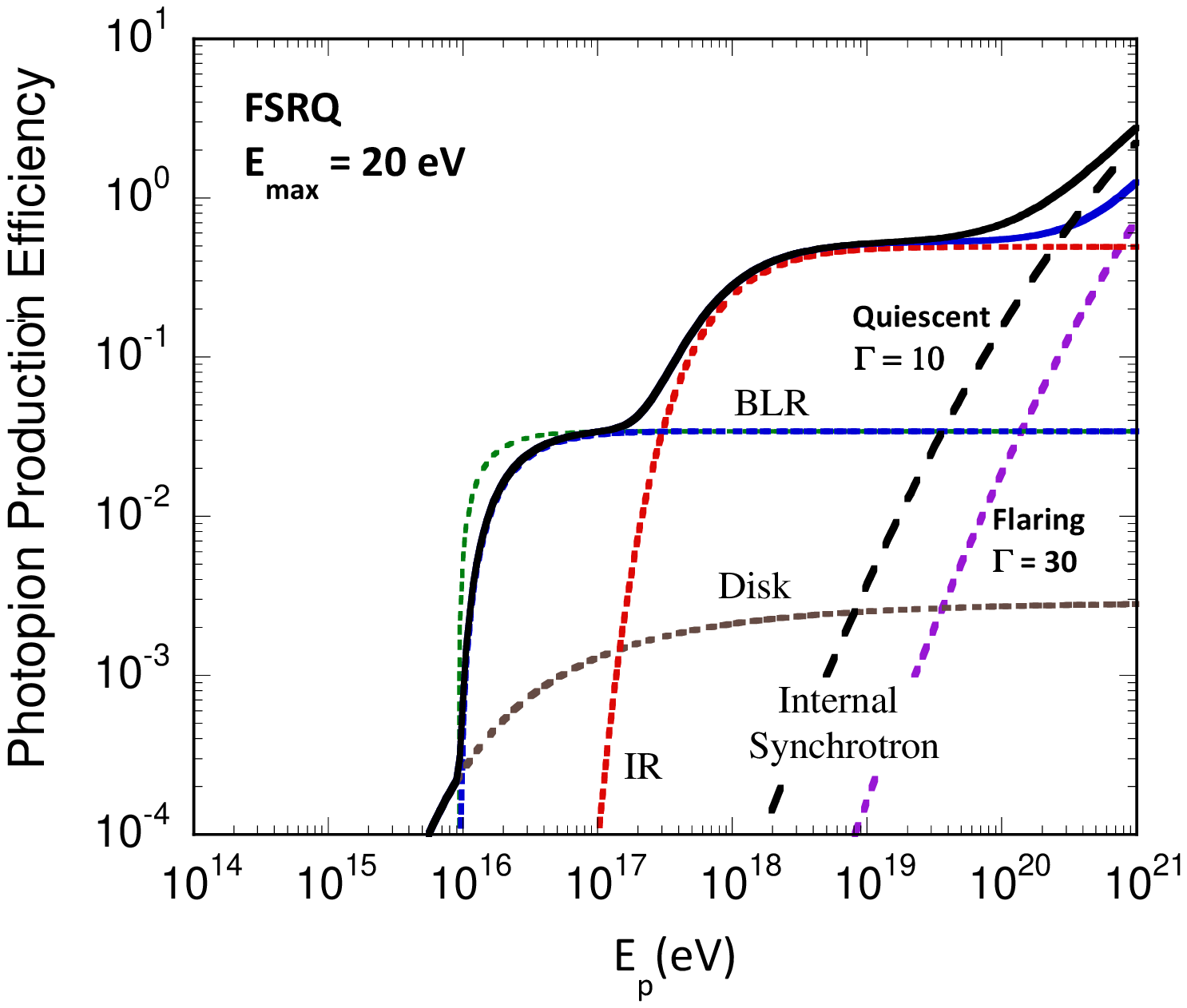}}
\vskip0.1in\raggedright 
\noindent{{\bf Fig.\ 3}
Minimum photopion production efficiency { as a function of escaping proton energy $E_p$} 
for parameters typical of quiescent and flaring states
of FSRQs. The efficiency for interactions with synchrotron radiation is 
determined by the dynamical timescale $t^\prime_{dyn} \cong  \Gamma t_{var}$, using values from Table 1,
 and $t^\prime_{dyn} = R_{ext}/(c\Gamma)$ for external processes, using values for 
the physical extent $R_{ext} = 0.1$ and $1$ pc of the BLR and IR radiation fields, respectively.
Separate contributions from photopion production with Ly $\alpha$ radiation in the BLR, scattered accretion-disk radiation,
IR radiation, and synchrotron photons are shown separately. The differing 
internal synchrotron efficiency for the quiescent and flaring cases are 
plotted by the long-dashed and short-dashed curves, respectively. Also plotted by the thin green dotted curve 
is the photopion efficiency for energy loss by a proton or neutron traveling rectilinearly through the
BLR radiation field. 
 }
 \label{Fig3}
\end{figure}

{ Photopion production efficiency $\eta_{\phi\pi}$ of high-energy protons with
 internal synchrotron photons is largest for small $\delta_{\rm D}$, because the photon density
is largest, so all injected power is reprocessed into neutrinos, $\gamma$ rays and neutrons. For internal processes,
the produced radiation can be assumed to be isotropically emitted in the comoving fluid frame, so
the neutrino luminosity $L_\nu \propto \delta^{4}L^\prime_\nu$, with the Jet Doppler opening angle decreasing $\propto
\delta^{-1}$. 
This leads to a characteristic 
Doppler factor $\hat \delta_{\rm D}$ when $\eta_{\phi\pi} \approx 1$ that optimizes neutrino luminosity; see Appendix B. }

\section{Photopion production efficiency in black-hole jet sources with external radiation fields.}

We now 
treat the case of black-hole jet sources with strong external radiation fields, most notably FSRQs, 
though LSP and ISP BL Lac objects with peak synchrotron frequencies $\lesssim 10^{15}$  Hz can also 
have external radiation fields with significant energy densities. In contrast, HSP BL Lac objects have radiatively inefficient accretion flows and generally lack evidence for optically thick accretion disks or luminous BLRs, so external radiation fields are usually neglected. As we have seen for equipartition values, and { as has been} shown earlier by detailed Monte Carlo simulations  \cite{mp01a}, blazars without external radiation fields radiate the bulk of the neutrinos' energy at $\gg 10^{17}$ eV, and would have difficulty explaining the IceCube PeV neutrinos { unless the Doppler factor was unusually low. As shown in Appendix B,  $\delta_{\rm D}\lesssim 4$ is required for lower-energy neutrino production in HSP BL objects, and would represent a system far from equipartition  with large $\gamma$-ray opacity that would produce absorption features that have not been observed in the SEDs of BL Lac objects   \citep{der14}.}

External radiation fields arise from accretion-disk radiation absorbed by and reradiated from the molecular torus and BLR clouds, and scattered by electrons (for recent reviews of AGN and blazar physics, see
\cite{bs12,bhk12}). The highly anisotropic direct accretion-disk
radiation field is shown in Appendix C to be unimportant for the production of PeV neutrinos.

{ The external radiation field from the accretion-disk radiation 
reprocessed by BLR clouds and the IR torus
is assumed to be have an isotropic distribution in the black-hole frame.
 Studies of anisotropies in the scattered radiation field \citep{der09,dp03}
show that locations within and close to the inner edge of the scattered radiation field
have approximately isotropic external fields. An increasing fraction of tail-on photons develop as
 the jet becomes closer to the outer edge of the scattering zone. Calculations of $\gamma$-ray and neutrino SEDs
entail a reaction-rate factor $(1-\beta\mu)$ that reduces the importance 
of tail-on photons. The assumption of  isotropy is a good first approximation well within the 
radiation reprocessing region, 
but should be relaxed in further studies.}

The transformation of an isotropic monochromatic external radiation field
with energy density $u_0$ and photon energy $\epsilon_0$ to 
the fluid frame is easily performed using the transformation 
law $u^\prime(\ep,\Omega^\prime) = u(\epsilon,\Omega)/[\Gamma(1+\beta\mu^{\prime})]^3$ for the 
specific spectral energy density $u(\epsilon,\Omega)$ (see eq.\ (5.24) in \citep{dm09}). For a highly
relativistic  ($\Gamma\gg 1$) flow, one obtains the spectral energy density 
$\ep u^\prime(\ep) \approx (u_0/2\Gamma)(\ep/\epsilon_0)^3 H(\ep;0,2\Gamma\epsilon_0)$,
after integrating over angle.
Substituting this expression into eq.\ (\ref{tprime-1}), noting eq.\ (\ref{nprimephep}) and multiplying by
$t^\prime_{dyn}$, gives the efficiency
\begin{equation}
\eta^{ext}_{\phi\pi} =\eta_0\,
[1 - {(1+\ln y_u)\over y_u}]H(y_u - 1)\;,\; \eta_0 \equiv {\hat\sigma u_0 R \over m_ec^2 \epsilon_0}\;,
\label{etaext}
\end{equation} 
where $y_u \equiv (4\epsilon_0 \gamma_p/\bar\epsilon_{thr})^2$ and the pathlength $R\lesssim R_{ext}$
through the target radiation field of extent $R_{ext}$. 
The comoving energy-loss rate for protons with escaping { energy $E_p = m_pc^2\gamma_p \cong m_pc^2\Gamma \gamma_p^\prime $ 
that lose energy through photopion processes with photons of a} locally isotropic external radiation fields is therefore given by
\begin{equation}
-\dot\gamma^\prime_{\phi\pi}(\gamma_p ) = {c\hat\sigma\gamma_p\over m_ec^2}\;\int_{\bar\epsilon_{thr}/4\gamma_p}^\infty d\epsilon\;{\epsilon u(\epsilon)\over \epsilon^2}\;[1 - {(1+\ln \bar y_u)\over \bar y_u}]\;,\;\;
\label{dgammaphipi}
\end{equation}
where $\bar y_u \equiv (4\epsilon \gamma_p/\bar\epsilon_{thr})^2$. In comparison with a proton bound in jet plasma moving with $\Gamma\gg 1$,
the corresponding efficiency of a neutron or proton traveling rectilinearly is
$\eta^{ext}_{\phi\pi} = \eta_0 (1-4/y_u)H(y_u - 4)$
\cite{mur12,footnote}.

We consider radiation fields associated with (1) the BLR, (2) the infrared-emitting dust torus, and (3) scattered accretion-disk photons,  
all of which provide target photons for { photopion production with cosmic rays coming from the jet}. In the first case, the Ly $\alpha$ radiation field dominates.
 For external isotropic monochromatic radiation, 
$\epsilon u_{0}(\epsilon)\approx \epsilon u_{0}\delta(\epsilon - \epsilon_0)$. 
In the specific case of  Ly $\alpha$ photons,
$\epsilon_0 = 2\times 10^{-5}$ is the Ly $\alpha$ photon energy 
in $m_ec^2$ units.
A spectrum of BLR lines 
has at most a small effect on the photon spectrum of Compton-scattered radiation \cite{cer13},
and similarly has a small effect on the neutrino spectrum except near the spectral cutoffs.
Nevertheless, we superpose a spectrum of lines in our subsequent neutrino production spectrum 
calculations.

For quasi-thermal infrared radiation from a dusty torus surrounding the black hole, 
$\e u_{IR}(\e ) =15 u_{IR}(\e/\Theta)^4/\{ \pi^4[\exp(-\e/\Theta)-1]\}$, where the 
effective IR temperature $T_{IR} = m_ec^2\Theta/k_{\rm B}$, and $u_{IR}$ is the 
energy density of the torus field, restricted by the blackbody limit
to $u_{IR} < u_{bb}(T) \cong 0.008 (T/1000\,{\rm K})^4$ erg cm$^{-3}$.

The third case involving scattered accretion-disk radiation is approximated by
 $\e u_{disk}(\e ) \approx u_{disk} (\e/\e_{max})^\alpha \exp (-\e/\e_{max} )$, where 
$u_{disk} = L_{disk}\tau_{sc}/\Gamma(\alpha) 4\pi R_{sc}^2 c$, $L_{disk}$ is the accretion-disk 
luminosity, and $\tau_{sc}$ is the Thomson depth through the scattering volume of radius $R_{sc}$. For a 
Shakura-Sunyaev spectrum, $\alpha = 4/3$, $\Gamma(4/3) = 0.893\dots$, and $\e_{max}$ corresponds
to the dimensionless temperature of the accretion disk near the innermost stable orbit, 
which must be $\gtrsim 2\times 10^{-5}$ in order to make strong Ly $\alpha$ radiation. 
In the calculations, we take $m_e c^2 \epsilon_{max} = 20$ eV.

Fig.\ 3 shows a calculation of the photopion production efficiency using typical 
parameters for $\gamma$-ray loud FSRQs. Compared to the sources in Fig.\ 2, the presence
of the external radiation field of the BLR, 
as well as the scattered accretion-disk radiation field, 
is extremely important for neutrino production in FSRQs \cite{ad01}. In this calculation, we take the energy density of 
the BLR radiation field $u_{BLR}= 0.026 (f_{BLR}/0.1)$ erg cm$^{-3}$ \citep{gt08}, 
where $f_{BLR}$ is the covering factor for atomic-line production. 
The BLR radiation is dominated by Ly $\alpha$,
but we also consider a range of lines with strengths given by analyses of AGN spectra \citep{cer13,tel02}, as given in Table 2. Furthermore, we 
assume that He Ly $\alpha$ lines are present with an energy density of one-half the Ly $\alpha$ energy density \cite{mid14,ps10}.
For the IR radiation field of the 
dust torus, we set  $u_{IR} = 10^{-3}$ erg cm$^{-3}$ and assume it has an effective temperature of 1200 K  \cite{mal11}. 

In addition, an electron column with effective Thomson scattering depth of $\tau_{sc} = 0.01$ in a 
region of extent $R_{sc} = 0.1$ pc is used in Fig.\ 3 to define the scattered accretion-disk radiation, which is approximated
by a Shakura-Sunyaev spectrum with temperature of 20 eV and $L_{disk} = 10^{46}$ erg s$^{-1}$. The direct accretion-disk radiation field provides another external photon 
target \cite{mp01}, but is unimportant for the production of PeV neutrinos (Appendix C), and is important for Compton 
scattering only if the emission region is within $\approx 10^{16}$ cm of the accretion disk \cite{ds02}. 

In the calculations of photopion efficiency,  $R$ is equated with $c \Gamma^2 t_{var}$ for
interactions with the internal radiation fields. 
For external radiation processes, where photopion production
can occur only as long as the jet remains within the target radiation field, 
the only requirement is that $R \lesssim R_{ext}$. 
For a BLR with $R_{ext}\sim 0.1$ pc, 
 a photopion production { efficiency $\eta_{\phi\pi}\approx 0.03$ can} 
be expected for $\gtrsim 10^{16}$ eV protons
in both the quiescent and flaring phases of FSRQs.

\begin{table}[t]
\begin{center}
\noindent\centering{\bf Table 2.} {BLR Emission Lines Included
in \\ the Modeling of Neutrino Production.$^a$}
\label{tablelines}
\begin{tabular}{c|c c}
\hline
\hline
Line & Flux & E (eV) \\
\hline
H Ly $\alpha$ & 100 & 10.2\\
C IV & 52.0 & 8.00 \\
He Ly $\alpha$ & 50.0 & 40.8\\
Broad feature$^b$ & 30.2 & 7.75 \\
Mg II & 22.3 & 4.43 \\
N V & 22.0 & 10.00\\
O VI + Ly $\beta$ & 19.1 & 12.04\\
C III + Si III & 13.2 & 6.53\\
\hline
\hline
\end{tabular}
\end{center}
\noindent{$^a$Line strengths are expressed as a ratio of the line flux to the H Ly $\alpha$ flux; see \citet{tel02,cer13}.\\
$^b$Broad feature at $\sim 1600\AA$ has equivalent width of $\approx 38.5\AA$ and is treated as a monochromatic line.}
\end{table}

\section{Neutrino production spectrum from photopion processes in black-hole jet sources}

Following  \cite{dm09,der12}, 
we derive the  neutrino luminosity spectrum for neutrinos made with energy of
m$_e$c$^2\epsilon_s$ by using a formalism where the photopion production cross section is divided into 
separate step functions. Here we consider a three step-function model, corresponding to single-,
double-, and multi-pion production that approximates the cross section and results from Monte Carlo 
simulations \cite{mue99,mue00}. Photopion interactions taking place with the invariant dimensionless
photon energy $\bar\epsilon_r$ in the range $\epsilon_{l,i}\leq \bar\epsilon_r < \epsilon_{u,i}$, $i, = 1,2,3$,
have cross section $\sigma_i$, neutrino multiplicity $\zeta_i$, and fractional energy $\chi_i$ of the neutrino
secondary (compared to the 
incident photon energy). The parameters for the model are given in Table 3, including $\beta$-decay neutrinos
from the decay of neutrons formed in photopion processes. Here we assume that neutrons are produced one-half of 
the time in photohadronic processes for single and double $\pi$ production, and one-third of the time for multi-pion production. This gives a rough approximation to the SOPHIA 2.0 event-generator neutron conversion efficiency \citep[see Fig.\ 11 in][]{mue00}.

\begin{table}[t]
\begin{center}
\noindent\centering{\bf Table 3.} {Parameters for secondary neutrinos 
formed in \\ photomeson production and neutron $\beta$ decay.$^a$}
\begin{tabular}{clll}
\hline
Parameter~~~ &  Single $\pi$~~~ &Double $\pi$~~~  &Multi $\pi$~~~  \\ 
i~~~ &  ~~1 & ~~2  &~~3  \\ 
\hline
\hline
$\sigma(\mu$b) & 340 & 180 &  120  \\
$\epsilon_l$ & 390 & 980 & 3200   \\
$\epsilon_u$ & 980 & 3200 &  $\infty$    \\
$\zeta$ & 3/2 & 4 & 6  \\
$\chi$ & 0.05 & 0.05 & 0.05   \\
$\zeta_\beta$ & 1/2 & 1/2 & 1/3  \\
$\chi_\beta$ & $10^{-3}$ & $10^{-3}$ & $10^{-3}$  \\
\hline
\hline
\label{table3}
\end{tabular}
\end{center}
\vskip-0.2in
\noindent $^a$ For neutron $\beta$-decay neutrinos, the same parameters 
as for photopion neutrinos are used except for multiplicity $\zeta_\beta$ and mean fractional energy $\chi_\beta$\\
\end{table}

For neutrino production from proton interactions with the internal synchrotron radiation field,
the synchrotron emission is assumed to be radiated by a distribution of electrons that are isotropically distributed in the comoving jet frame and described by a log-parabola
function \citep{der14}. 
The synchrotron luminosity spectrum is given by eq.\ (\ref{eLsyne}), and the 
synchrotron photon spectrum coming from relativistic electrons 
is  given by eq.\ (\ref{nprimephep}).
 The photohadronic  production cross section 
for secondary neutrinos is approximated by
$${d\sigma(\bar\epsilon_r)\over d\epsilon_s^\prime d\Omega_s^\prime} =$$
\vskip-0.2in
\begin{equation}
 \sum_{i=1}^3 \zeta_i \sigma_i H(\bar\epsilon_r; \epsilon_{l,i},\epsilon_{u,i})
\delta(\Omega^\prime_s -\Omega^\prime_p)\delta(\epsilon_s^\prime - {\chi_i m_p\gamma_p^\prime\over m_e})\;,
\label{nprimeepsilonprime}
\end{equation}
making the co-directional approximation that the secondaries travel in the same direction as the primary ultra-relativistic
proton, and that the secondary energy is a fixed fraction $\chi$ of the primary energy. 
Here $\bar\epsilon_r = \gamma_p^\prime \epsilon^\prime (1-\mu^\prime)$ is the invariant collision energy, 
and $H(x;a_1,b_1)= 1$ if $a_1 \leq x \leq b_1$, and $H(x;a_1,b_1)= 0$ otherwise.

For the description of the proton spectrum in the blob, we also adopt the log-parabola function, and assume for simplicity that the log-parabola width parameter $b$ is the same for protons as electrons (differing from the treatment in Ref.\ \cite{mid14}). The spectrum of protons with Lorentz factor $\gamma_p = \delta_{\rm D} \gamma_p^\prime$ is therefore given by
\begin{equation}
\gamma_p^{\prime 2} N_p^\prime(\gamma_p^\prime ) = K x_p^{-\hat b \ln x_p}\;,
\label{gammap2}
\end{equation}
where $K\equiv {\cal E}_p^\prime/m_pc^2 I_1(b)$, $x_p = \gamma_p/\gamma_{pk} = \gamma^\prime_p/\gamma^\prime_{pk}$, ${\cal E}_p^\prime$ is the total comoving energy of the nonthermal protons, and 
$I_1(b) = \sqrt{\pi\ln 10/b}$ \cite{der14}.
Because $\epsilon_s L(\epsilon_s,\Omega_s) = \delta_{\rm D}^4 \epsilon^\prime_s L^\prime(\epsilon^\prime_s,\Omega^\prime_s)$, 
one obtains
$$4\pi\epsilon_s L^{int}(\epsilon_s,\Omega_s) = \sum_{i= 1}^3{K\zeta_i m_e \sigma_i \epsilon_s^2 \upsilon  \tilde x^{-4-\hat b \ln \tilde x}
 \over 16\pi  f_0 \chi_i m_p c^2 t_{var}^2 }$$
\vskip-0.2in
\begin{equation} 
\times \int^\infty_{\epsilon_{l,i}}d\epsilon^\prime {y^{1-\hat b \ln y}\over \epsilon^{\prime 4}}\,
\{ [\min(\epsilon_{u,i},2\tilde\gamma_p^\prime \epsilon^\prime)]^2 - \epsilon_{l,i}^2 \}
\label{esLesint}
\end{equation}
for the neutrino production spectrum from photohadronic interactions with synchrotron photons.
Here $\tilde x = \tilde \gamma_p/\gamma_{pk}$, $ \tilde \gamma_p = m_e \epsilon_s/\chi_i m_p $, and 
$\tilde \gamma_p^\prime = \tilde \gamma_p/\delta_{\rm D}$.

We follow the technique of Ref.\ \cite{geo01} to derive the production spectrum of neutrinos 
formed when protons interact with photons of an external 
isotropic radiation field, by transforming the particle distribution to the source frame directly
(see also \cite{der12}). The result is
$$4\pi\epsilon_s L^{ext}(\epsilon_s,\Omega_s) = \sum_{i=1}^3{K\zeta_i m_e \sigma_i c \delta_{\rm D}^5 \epsilon_s^2  \tilde x^{-4-\hat b \ln \tilde x}
 \over 4  \chi_i m_p \gamma_{pk}^4 }$$
\vskip-0.2in
\begin{equation} 
\times \int^\infty_0 \,d\epsilon {u(\epsilon)\over \epsilon^{3}}\,
\{ [\min(\epsilon_{u,i},2\tilde\gamma_p \epsilon^\prime)]^2 - \epsilon_{l,i}^2 \}\;.
\label{esLesext}
\end{equation}
Note the $\delta_{\rm D}^5$ dependence \citep{der12}. The $\delta$-function approximation
to the neutrino production spectrum does not give a good representation to the low-energy cutoff
of the neutrino spectrum, which follows a number spectral index of $-1$ \cite{ste79}. 
For pion-decay neutrinos formed with target synchrotron, BLR, scattered accretion-disk 
and IR photons,  we improve the approximation
by correcting the neutrino spectrum by adding a low-energy extension with $\nu F_\nu$ index
 equal to $+1$ if the
$\nu F_\nu$ spectrum calculated in the $\delta$-function approximation to 
the mean neutrino energy becomes harder than $+1$. 
No correction is made for the spectrum of $\beta$-decay neutrinos in the $\delta$-function approximation for
 average neutrino energy. For detailed numerical calculations, see, e.g., Ref.\ \cite{tak09}.

\begin{figure}[t]
\centering
{\includegraphics[scale=0.5]{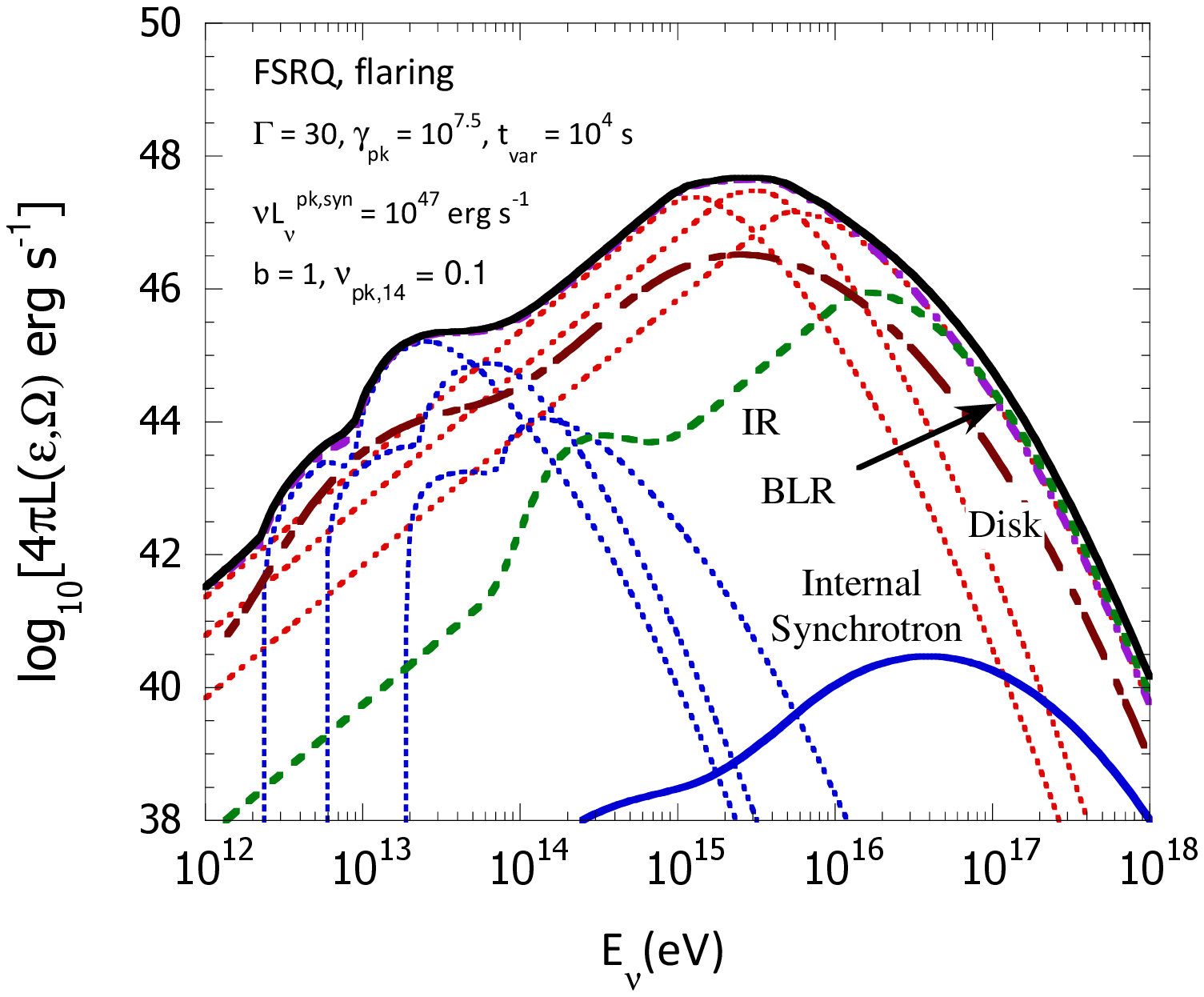}}
\vskip0.1in\raggedright
\noindent{{\bf Fig.\ 4}
The  luminosity spectrum of neutrinos of all flavors 
from an FSRQ with 
$\delta_{\rm D} = \Gamma = 30$, 
using parameters of a flaring blazar given in Table 1.
The  radiation fields are assumed isotropic 
with energy densities  $u_{BLR} = 0.026$ erg cm$^{-3}$ for the BLR field,  $u_{IR} = 0.001$ erg cm$^{-3}$ for
the graybody IR field. For the scattered accretion-disk field,  $\tau_{sc} = 0.01$ is assumed. The proton spectrum is described by a log-parabola function with 
log-parabola width $b=1$ and principal Lorentz factor $\gamma_{pk} = \Gamma \gamma^\prime_{pk} = 10^{7.5}$.
Separate single-, double- and multi-pion components comprising the neutrino luminosity spectrum 
produced by the BLR field are shown by the light dotted curves for the photohadronic and $\beta$-decay
neutrinos. Separate components of the neutrino spectra from photohadronic interactions with the synchrotron, 
BLR, IR, and scattered accretion-disk radiation are labeled.
}
 \label{Fig4}
\end{figure}

Fig.\ 4 shows a calculation of the luminosity spectrum of neutrinos of 
all flavors produced by a curving distribution of protons in a flaring FSRQ like 3C 279 with a peak synchrotron 
frequency of $10^{13}$ Hz and peak synchrotron luminosity of $10^{47}$ erg s$^{-1}$ (parameters of 
Table 1). The log-parabola width
parameter $b = 1$ is assumed for both the electron and proton distributions. Here and below,
we take ${\cal E}_p^\prime = 10^{51}/\Gamma$ erg, 
which implies sub-Eddington jet powers for jet ejections occurring no more frequently than once every $10^{4}M_9$ s, 
where $M_9$  is the black-hole mass in units of $10^9$ M$_\odot$ (we take $M_9 = 1$). 
The separate components for 
single-pion, double-pion, and multi-pion production from interactions with the 
BLR radiation are shown for both the pion-decay and neutron $\beta$-decay neutrinos. 
In this calculation, the proton principal Lorentz factor $\gamma_{pk} = 10^{7.5}$, 
corresponding to source-frame principal proton energies of $E_p \approx 3\times 10^{16}$ eV.
Because the efficiency for synchrotron interactions in  low-synchrotron peaked blazars
is low until $E_p \gtrsim 10^{20}$ eV, as seen in Fig.\ 3, neutrino production from the synchrotron component is 
consequently very small.  Interactions with the blazar BLR radiation is most important, resulting for this value of $\gamma_{pk}$ in a neutrino luminosity 
spectrum peaked at a few PeV, and with a cutoff below $\approx 1$ PeV.

\begin{figure}[t]
\centering
{\includegraphics[scale=0.5]{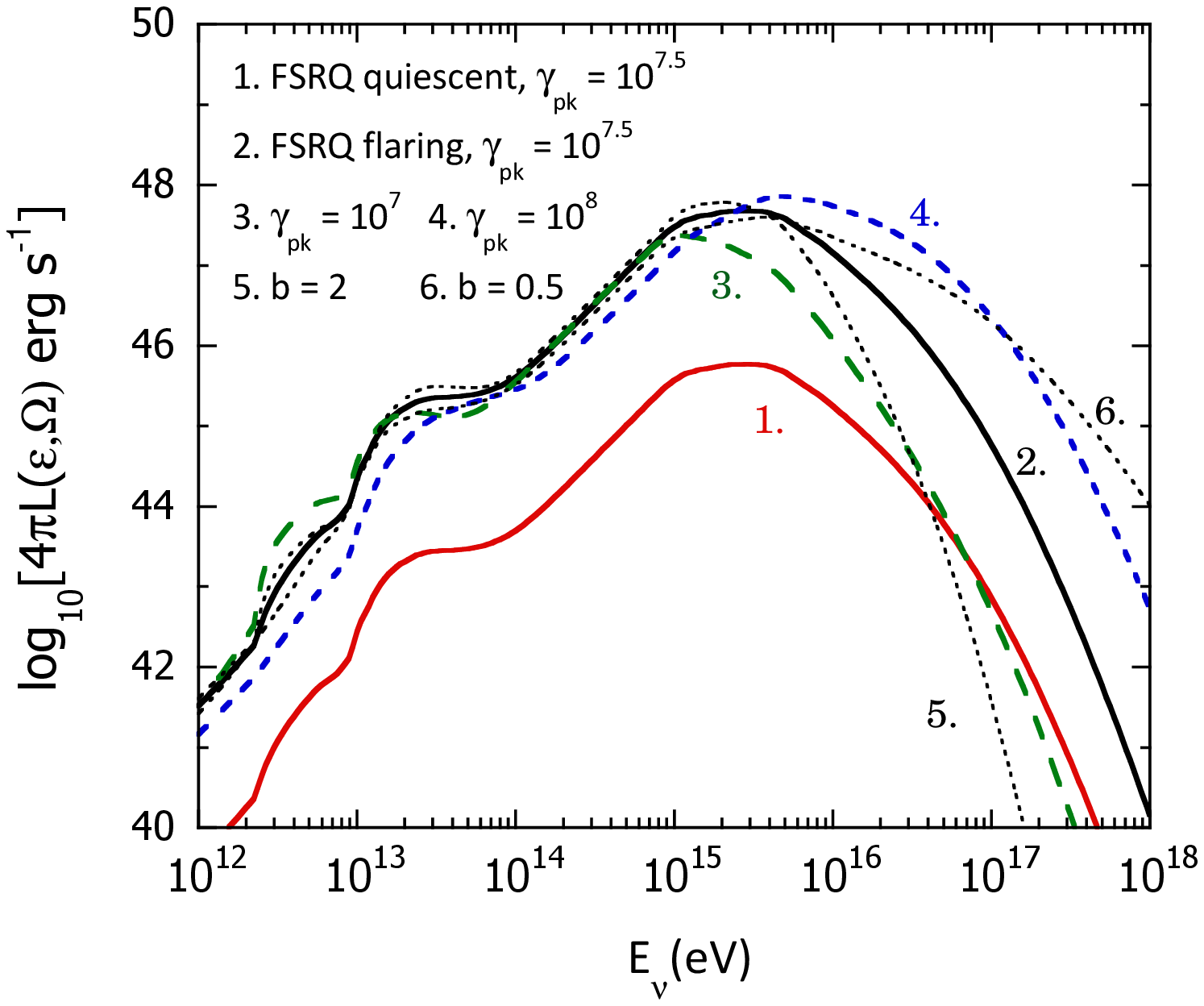}}
\vskip0.1in \raggedright
\noindent{{\bf Fig.\ 5}
Total luminosity spectra of neutrinos of all flavors from model FSRQs with parameters as given 
in Fig.\ 4, except as noted. In curve 1, parameters of a quiescent blazar from Table 1, with $\gamma_{pk} = 10^{7.5}$, are used. Curves 2 -- 6
use parameters for a flaring blazar as given in Table 1. In curves 2, 3, and 4, $\gamma_{pk} = 10^{7.5}$, $10^7$, and 
$10^8$,  respectively. Curves 5 and 6 use the same parameters as curve 2, except that 
 $b = 2$ and $b = 0.5$, respectively.
}
 \label{Fig5}
\end{figure}

Comparisons between luminosity spectra of neutrinos of all flavors for parameters corresponding to the 
quiescent phase of blazars, and for different values of $\gamma_{pk}$ and $b$, as labeled, are shown
in Fig.\ 5. As can be seen, the low-energy hardening in the neutrino spectrum below $\approx 1$ PeV is insensitive to the assumed values of $\gamma_{pk}$ and $b$.

\section{Discussion}
We have calculated the efficiency of neutrinos produced
 by photohadronic interactions of protons with internal and external target photons
in black-hole jet sources. 
{  Neutrino spectra were calculated semi-analytically for the
chosen parameters. 
After summarizing (1) data from IceCube motivating this study, we discuss 
 (2) the UHECR/neutrino connection, 
(3) particle acceleration in jets, 
and (4) the contributions of FSRQs and blazars to the diffuse neutrino background.

\subsection{Extragalactic Neutrinos with IceCube}

The IceCube Collaboration has reported compelling
evidence for the first detection of high-energy neutrinos 
from extragalactic sources.
The sources of the neutrinos remain unknown.
Candidate astrophysical sources include powerful 
$\gamma$-ray sources such as blazars, GRBs, and young pulsars or
magnetars. Other possibilities, e.g.,
structure formation shocks and star-forming galaxies, are 
not excluded. Here we have argued 
that FSRQs are $\gtrsim 1$ PeV 
neutrino sources.

IceCube searches have not, however, 
found statistically compelling counterparts
by correlating neutrino arrival directions and times with 
pre-selected lists of candidate neutrino point sources, including
FSRQs. An early
search  \citep{abb09} using 22-string data over 276 days live time
found no significant excess other than 1 event associated with PKS 1622-297.
Upper limits for an $E^{-2}$ neutrino spectrum from candidate $\gamma$-ray emitting AGNs 
were at the level of $\approx 
1.6\times 10^{-12}\Phi_{90}$ erg cm$^{-2}$ s$^{-1}$, $15\lesssim \Phi_{90} \lesssim 600$,
 for neutrinos with 
energies $E_\nu$ from $\approx 100$ TeV to $\approx 100$ PeV. The upper limit for 3C 279
was a factor $\gtrsim 30$ above model predictions \cite{rei09,ad01}.

Improved point-source searches in 22-string and 40-string configurations during 
2007 -- 2009 were reported for both flaring and persistent sources 
in Ref.\ \citep{abb12}.
Recent 86-string data taken over 1373 days live time give IceCube limits
of $\approx  10^{-12}$ erg cm$^{-2}$ s$^{-1}$ for 1 TeV $\lesssim E_\nu \lesssim 1$ PeV
in the northern sky, and $\approx  10^{-11}$ erg cm$^{-2}$ s$^{-1}$ for 
100 TeV $\lesssim E_\nu \lesssim 100$ PeV
in the southern sky  \citep{aar14a}.

Source $\gamma$-ray fluxes provide an upper limit to the neutrino flux because
the decay of $\pi^0$ and $\pi^\pm$ formed in photopion process will produce
secondaries that initiate $\gamma$-ray cascades that cannot overproduce the measured $\gamma$-ray 
fluxes. The brightest $\gamma$-ray blazars, namely 3C 279, 3C 273, and 3C 454.3,
have average $> 100$ MeV fluxes
at the level of $\approx$ few $\times 10^{-11}$ erg cm$^{-2}$ s$^{-1}$ \citep{abd09}.
These limits rule out a hypothetical blazar model where the $\gamma$ rays are entirely
associated with photohadronic processes, but the success of leptonic models for blazar
$\gamma$ radiation \citep{bhk12} means that only a small fraction of the high-energy radiation
from blazars can be hadronically induced. Particular interest for neutrino counterpart association
attaches to unusual very-high energy (VHE; $\gtrsim 100$ GeV) flaring episodes in FSRQs, 
such as 3C 279 \citep{alb08} and PKS 1222+216 \citep{ale11}. 
Furthermore, analysis of associations between GeV-TeV sources and IceCube neutrino arrival directions 
finds counterpart TeV BL Lac objects and pulsar wind nebulae 
 \citep{pr14}.  In principle, two-zone models for
these objects could achieve the required flux \citep{tgg14} by adjusting the
cosmic-ray spectral index and cutoff energy to appropriate values, but
one has to take into account contributions from FSRQs for a detailed
comparison.

}

\subsection{UHECR/High-Energy Neutrino Connection}

High-energy neutrino sources are obvious UHECR source candidates, though production of 
PeV neutrinos requires protons with energies of ``only"  $E_p \cong 10^{16}$ -- $10^{17}$ eV.
The close connection between neutrino and UHECR production implies the well-known Waxman-Bahcall 
(WB) bound on the diffuse neutrino intensity at the level of $\sim 3\times 10^{-8}$ GeV/cm$^2$-s-sr \cite{wb99}, 
and the similarity of the IceCube PeV neutrino flux with the WB bound has been noted \cite{wax13}.
Nevertheless, our results show that the  relationship between the diffuse neutrino and UHECR intensities leading to the WB bound depends on fine-tuning the neutrino production and escape probability of UHECRs.  The situation is even worse if the UHECRs are ions rather than protons, because  the photo-disintegration cross section for ions is larger than the photohadronic cross section and neutrino production is less efficient \cite{wan08,mur08}. 

For GRBs and HSP BL Lac objects, the most significant radiation field for photopion production is the internal synchrotron field. { Appendix B gives the Doppler factor for optimal neutrino production, $\hat \delta_{\rm D}$, and the typical energy $E_\nu$ of the produced neutrinos in 
terms of the apparent isotropic synchrotron luminosity, the peak synchrotron frequency, and the minimum variability time. As a consequence of the low value of the peak synchrotron frequency, FSRQs formally require $\delta_{\rm D} \sim 70$ to make $\sim 10^{20}$ eV neutrinos, but would have to accelerate protons to $\gtrsim 10^{21}$ eV. GRBs and BL Lac objects with small Doppler factors can effectively make $\sim 100$ TeV neutrinos from photopion losses on internal synchrotron photons. The SEDs in one-zone models of such low-Doppler factor BL Lac objects would, however, be strongly distorted by internal $\gamma\gamma$ absorption. Furthermore,}
provided that $t_{var}$ and $\Gamma$ (or $\delta_{\rm D}$) are sufficiently small so that the internal target photon density is large (eq.\ (\ref{etas})), efficient photopion and neutrino production can take place in  GRBs, including LLGRBs. GRBs are also extremely powerful, so can accelerate protons to $\gtrsim 10^{20}$ eV
from simple arguments regarding Fermi acceleration (e.g., \cite{wax95}). Except under special conditions that $\eta_{\phi\pi} \sim 1$ is realized for typical $\Gamma$, however, GRBs would be weak neutrino and strong UHECR sources when $\Gamma$ is large, and strong neutrino sources with quenched UHECR production when $\Gamma$ is small (Fig.\ 2).

This difficulty also exists for blazars. 
HSP BL Lac objects are always inefficient PeV neutrino producers for the assumed parameters, 
as seen in Fig.\ 2, { and when they are efficient PeV neutrino emitters, $\gamma$-ray 
opacity is large, contrary to the appearance of the $\gamma$-ray SEDs of these objects.}
Because their SEDs are well described by nonthermal 
synchrotron self-Compton models, values of Doppler factor and fluid
magnetic field can be determined, { which are similar to values found
in equipartition modeling \citep{der14}}. Using such values from spectral modeling, 
along with the \citet{hil84} condition to define the maximum possible particle energy, 
Ref.\  \citep{mur12} found that HSP BL Lacs are not capable of accelerating protons to $E_p\gtrsim 10^{19}$ eV. 
If BL Lac objects are the sources of the UHECRs, then a transition from 
light to heavy composition would be required, as indicated in Auger \cite{abr10} (though not HiRes; \cite{abb05}) analyses of UHECRs. 
Indeed, BL Lac objects and their off-axis counterparts (i.e., FRI radio
galaxies) may be favored as sources of UHECRs because they are found
within the GZK radius, and their $\gamma$-ray emissivity greatly exceeds the UHECR
emissivity \cite{dr10}. If the UHECR source spectrum has a log-parabolic type behavior,
then the second-knee and ankle structures in the cosmic-ray spectrum, 
as well as compositional changes, could result from a superposition of UHECR injection spectra
modified by transport and energy losses, just as it is for power-law injection. 
Fits to the UHECR spectrum from blazar sources is, however, beyond the scope of the 
present work. 

Escaping UHECRs from the jets of BL Lac objects
can explain various peculiarities in blazar physics, including the hardening of 
the deabsorbed TeV spectrum for most models of the extragalactic background light (EBL), 
and the existence of an unusual, weakly variable class of TeV blazars \citep{ek10,ess11,ek12}. 
Indeed, production of neutrinos formed by very high energy cosmic-ray protons from a blazar source 
in transit through the intergalactic medium
has been proposed as an explanation for the PeV events \cite{kal13} 
based on calculations made prior to the detections \cite{ess10}.
The model as proposed cannot, however, explain
PeV neutrinos and UHECRs simultaneously. This is 
because the maximum cosmic-ray energy has to be tuned to $\lesssim 10^{17.5}$ eV
in order not to overproduce multi-PeV neutrinos. In addition, 
high EBL models, which are challenged by GRB observations \cite{abd10}, are needed.
Moreover, the neutrino spectrum must 
harden below $\approx 1$ PeV because the EBL is cutoff above $\approx 13.6$ eV. 
This model therefore needs other components such as star-forming 
galaxies and galaxy clusters to explain sub-PeV neutrino events.

Neutrino production from proton interactions in the inner jets of FSRQs differs significantly from 
the preceding types of sources by virtue of the strong external radiation fields that are required 
when modeling their $\gamma$-ray SEDs.  Indeed, FSRQs are defined by the strength of their broad lines. Though leptonic models appear adequate to explain the broadband SEDs of FSRQs, a hadronic component can explain observations of VHE $\gamma$ rays in FSRQs \cite{boe13}.
The calculations presented here show that if high-energy cosmic-ray protons are accelerated
in the inner jets of FSRQs, photopion losses with $\approx 1$ -- 10\% efficiency is 
found for both FSRQs in their quiescent and flaring states, but that the proton spectrum must soften at $E_p \gtrsim 10^{16}$ eV due to the assumed log-parabolic function, and can therefore not be significant UHECR sources. The dominant radiation field is the BLR radiation, though scattered accretion-disk radiation and, at higher proton energies, IR radiation, can also result in efficient photopion losses.

Figs.\ 4 and 5 show that a distinct low-energy hardening in the neutrino spectrum below 1 PeV is formed, as explained in the Introduction. Compared to the sharp cutoff found for a single monochromatic external radiation field, some smoothing is formed by a distribution of target photons from atomic lines, a stronger scattered accretion-disk radiation field, and a low-energy extension of the neutrino number spectrum $\propto E_\nu^{-1}$. Even the inclusion of a distribution in redshifts $z$ of FSRQs in the calculation of the diffuse neutrino flux from blazars (see Figs.\ 13 -- 16 in \cite{mid14}), which range broadly from $z\approx 0.5$ to  $z \approx 2$ \citep{ack11}, is not sufficient to remove this low-energy hardening, which appears to be a robust feature of FSRQs. If the hardening is not found in IceCube data, then other sources must be considered to fill in the gap and explain LE neutrinos. 

If the spectra of cosmic rays in star-forming galaxies are like our Galaxy's cosmic-ray spectrum, with the cosmic-ray proton spectrum softening at the knee ($\approx 3$ PeV), then the neutrino-production spectrum through secondary nuclear processes should soften at $\sim 0.05\times 3$ PeV, or at $\approx 150$ TeV. The LE neutrinos could then be due to 
superposition of neutrino emissions from star-forming/starburst galaxies or galaxy clusters
and groups. A higher energy cutoff in the proton spectrum of star-forming galaxies or galaxy clusters and groups may consistently explain the PeV neutrinos if the cosmic-ray number index is
 $\lesssim 2.1$ -- 2.2, in order to avoid overproducing the extragalactic $\gamma$-ray background \citep{mur13}.
Such sources of $\approx 0.03$ -- 2 PeV neutrinos require cosmic rays of $\approx 0.6$ -- 80 PeV energy (for a typical redshift $z\approx 1$).
Furthermore, neutrino emission from nuclear collisions in the Fermi bubbles might explain some though not all 
of the LE neutrinos \citep{lun13,am13}.


If scattered accretion-disk radiation is the dominant radiation field for photopion production, then a cutoff below $\sim 10^{14}$ eV can result if the accretion-disk radiation has an effective temperature of $\approx 35$ eV  \cite{ad01} (the mean photon energy is $\approx 3\times$ the temperature). This way to make LE neutrinos is, however, problematic by requiring unusually large ($\gg 10$ eV) effective temperatures for the accretion disks in FSRQs and large scattering depths $\tau_{sc}\sim 1$, leading to an associated $\gamma\gamma$ opacity that strongly attenuates the blazar $\gamma$-ray spectrum down to a few GeV. Neutrino production from the cores of AGNs, without distinguishing radio-loud and radio-quiet sub-classes, was proposed in Ref.\ \cite{ste91}. As originally formulated, this model overproduces the IceCube neutrino intensity by a large factor, but could have the observed flux level by renormalizing the injected cosmic-ray flux  \cite{ste13}. 

To summarize, given large baryon loading \cite{mid14}, 
FSRQs represent viable sources of the IceCube PeV neutrinos, but
make a  hardening below $\sim 1$ PeV, so that the neutrinos with $E_\nu \ll 1$ PeV
have to be made by another neutrino source class. Furthermore, 
the cosmic-ray energy distributions in FSRQs must
have a high-energy softening, which we have modeled with a log-parabola
spectrum, meaning that FSRQs cannot be the sources of the UHECRs. For
this, HSP BL Lac objects are favored in terms of emissivity and existence of
sources within the GZK radius, provided that their particle distribution
extends, unlike in FSRQs, to ultra-high energies. 


\subsection{Cosmic-ray acceleration in black-hole jets}

A log-parabola function has been assumed for the proton spectrum, which is a departure from power-law 
particle spectra that are usually assumed.  The mechanisms accelerating particles in 
blazars and the prompt-phase emissions of GRBs are highly uncertain. 
A curving log-parabola function can often give a better fit to the blazar SED, with fewer parameters, than electron spectra formed by the injection of power-laws followed by adiabatic losses and radiative cooling \cite{cer13,der14}. In both
blazars and prompt emissions of GRBs, the synchrotron radiation spectrum never reaches its maximum energy 
of $\approx 100 \Gamma$ MeV, so that a slower, second-order acceleration scenario that results
in curving particle distributions may be favored. (The delayed onset of $\gamma$-ray emissions at GeV energies
could, however, be synchrotron radiation from first-order Fermi acceleration of electrons at the external blast-wave shock \cite{kbd09,ghi10}.)
A curving proton distribution, or a soft power-law distribution, is consistent with the lack of a 
large flux of multi-PeV neutrinos. Nonlinear
effects in first-order acceleration make concave particle spectra, opposite to the behavior required to 
explain the IceCube data. On the other hand, a long acceleration timescale compared to escape could cause a cutoff at high energies in the particle spectra formed in first-order Fermi acceleration.

The simplest characterization of the maximum particle energy is to suppose that the relevant mechanism is Fermi acceleration, which operates on timescales longer than the Larmor time $t^\prime_{\rm L} = E^\prime/(QB^\prime c)$, where $E = \Gamma E^\prime$ is the escaping particle energy and $Q = Ze$ is its charge. For first-order Fermi acceleration, this implies a characteristic timescale $t^\prime_{\rm F1} = f_1 t^\prime_{\rm L}$, with $f_1 \gtrsim 1$. If the dynamical timescale $t^\prime_{dyn}$ during which the accelerator is active is determined by the measured variability timescale $t_{var}$, then $t^\prime_{dyn}\cong \Gamma t_{var}\cong \delta_{\rm D}
t_{var}$, and
\begin{equation} 
{t^\prime_{\rm F1}\over t^\prime_{dyn}} \cong f_1 \left( {E\over QB^\prime c\Gamma^2 t_{var}}\right)\;.
\end{equation} 
The condition ${t^\prime_{\rm F1}/ t^\prime_{dyn}}\cong 1$, with $f_1 \cong 1$, is a restatement of the \citet{hil84} condition that gives the maximum energy $E$ of a particle with charge $Q$. 

Using simple forms for particle acceleration derived in Ref.\ \cite{dml96} for gyroresonant acceleration of protons with Alfv\'enic turbulence, the corresponding relation for second-order Fermi acceleration is 
\begin{equation} 
{t^\prime_{\rm F2}\over t^\prime_{dyn}} \cong f_2 \left( {E\over QB^\prime c\Gamma^2 t_{var}}\right)^{2-q}\;.
\end{equation} 
Here $q$ is the index of turbulence, with $q = 5/3$ for Kolmogorov turbulence and $q = 3/2$ for Kraichnan turbulence, and the term $f_2 \equiv 2q/[\pi (q-1) \beta_{\rm A}^2 \zeta]$, where $\zeta$ is the fraction of magnetic-field energy density in Alfv\'enic turblence, and $c\beta_{\rm A}$ is the Alfv\'en speed.

\begin{figure}[t]
\centering
{\includegraphics[scale=0.5]{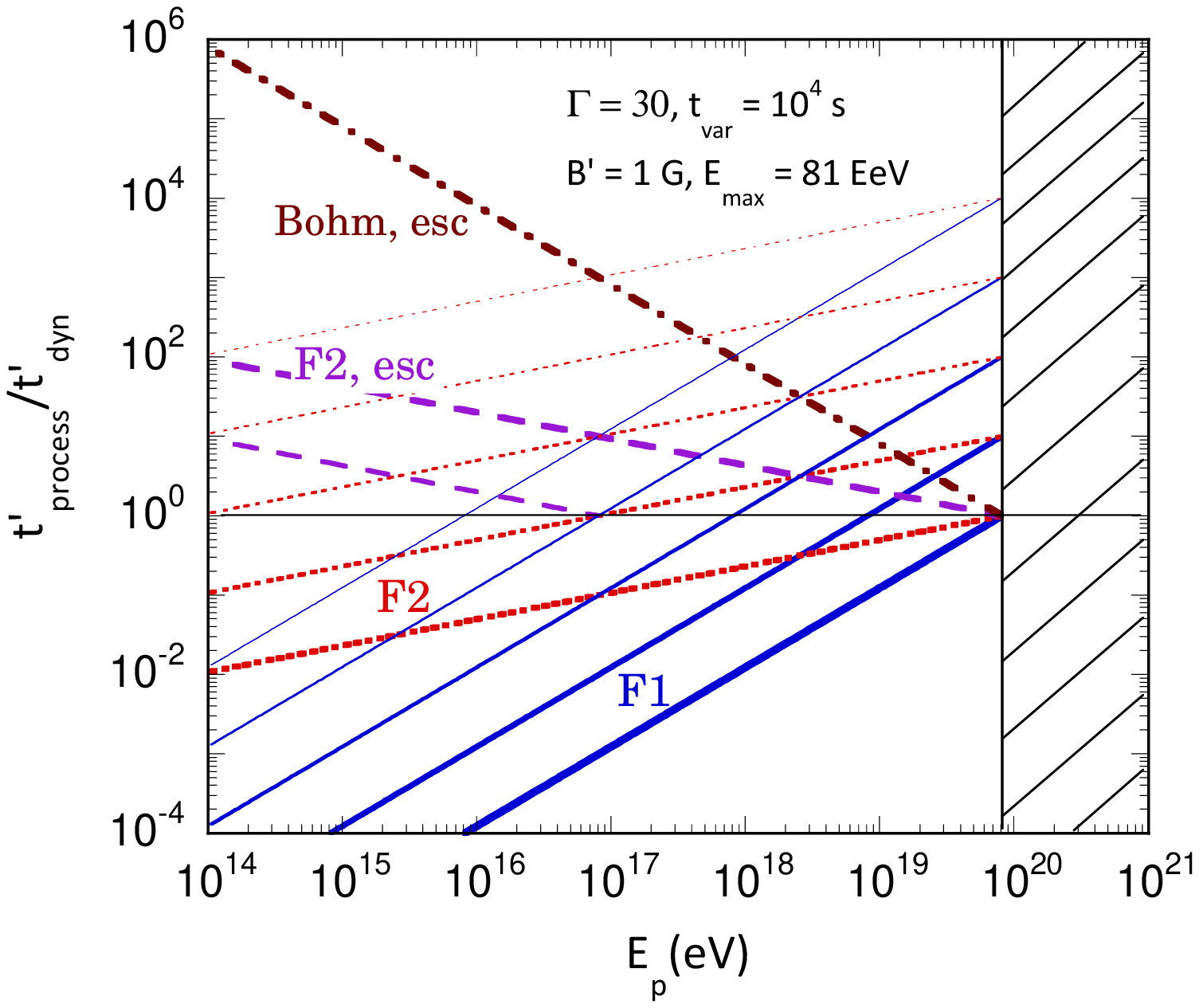}}
\vskip0.01in \raggedright
\noindent{{\bf Fig.\ 6}
Ratios of proton acceleration and and escape timescales to the dynamical timescale in the fluid frame are plotted { as 
a function of escaping proton energy $E_p\cong \Gamma E_p^\prime$} for FSRQ parameters given in the legend.  Acceleration efficiency reaches a maximum for both simplified descriptions of first-order (F1) and second-order (F2) Fermi acceleration, shown by heavy solid and dotted lines, respectively, with progressively lighter lines corresponding
to a reduction in the acceleration efficiency by an order-of-magnitude. Ratios of maximum escape times to the dynamical time through diffusive gyroresonant pitch-angle scattering and Bohm diffusion are shown by the labels ``F2,esc" and ``Bohm,esc," respectively. Calculations for second-order Fermi acceleration assume $q=5/3$. For the chosen parameters, protons cannot be accelerated to energies found in the cross-hatched region according to the Hillas criterion.
}
 \label{Fig6}
\end{figure}

\begin{figure}[t]
\centering
{\includegraphics[scale=0.5]{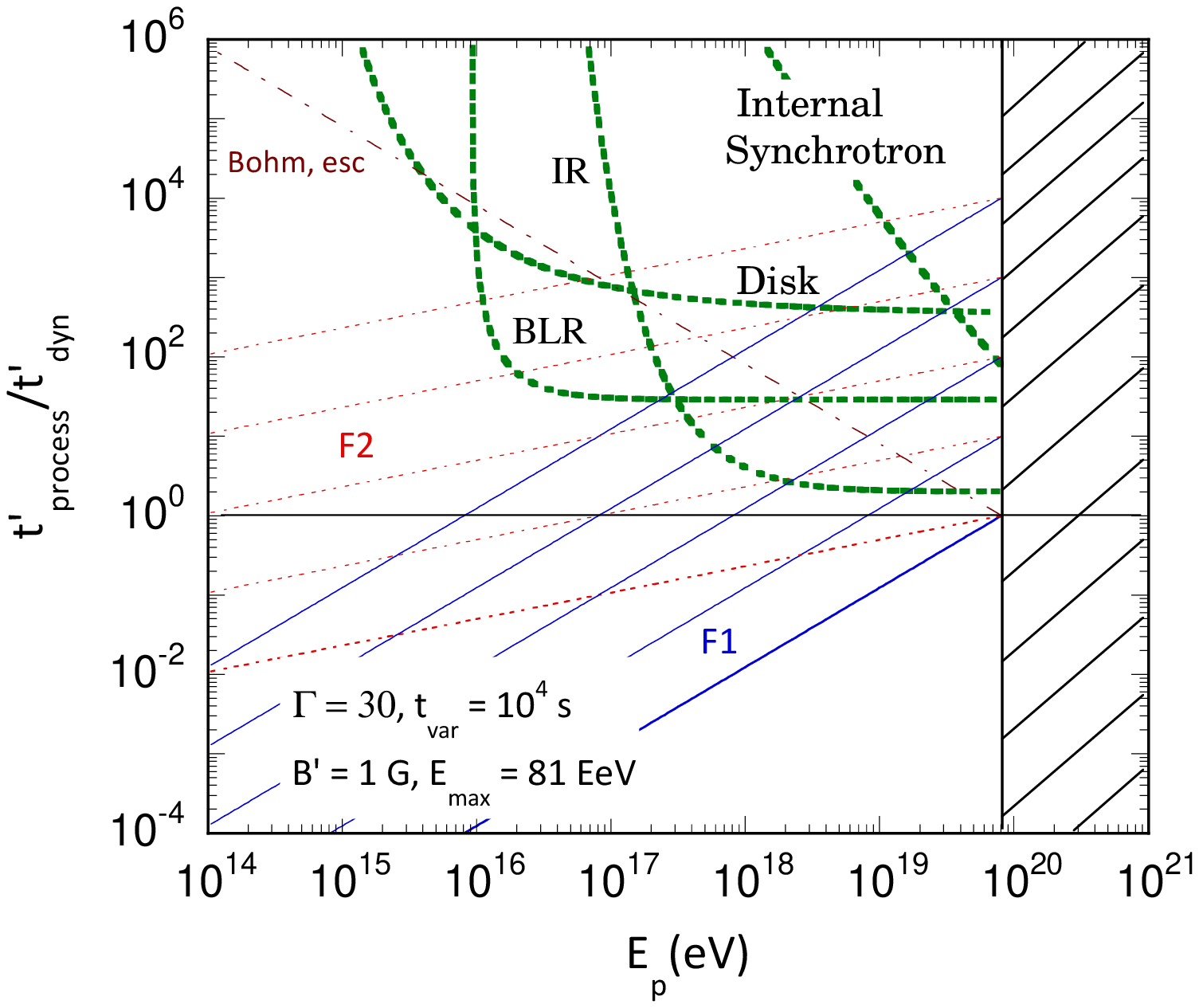}}
\vskip0.01in \raggedright
\noindent{{\bf Fig.\ 7}
Ratio of comoving proton photopion energy-loss timescale to the dynamical timescale are plotted for parameters of a flaring FSRQ  given in the legend. The BLR and IR have energy densities $u_{BLR} = 0.026 $ erg cm$^{-3}$ and $u_{IR} = 10^{-3}$ erg cm$^{-3}$, respectively, and the scattered disk component represents a scattering shell of Thomson depth $\tau_{\rm T} = 0.01$ and an optically thick accretion disk with an effective temperature of 20 eV.
}
 \label{Fig7}
\end{figure}

In Fig.\ 6, we plot the ratios of the comoving particle acceleration, energy-loss, and escape timescales to the dynamical timescale $t^\prime_{dyn}\cong \Gamma t_{var}$. Parameters are appropriate to a model flaring blazar, and are given in the figure legend. For second-order processes, a Kolmogorov turbulence spectrum ($q = 5/3$) is assumed. The characteristic times to accelerate protons to energies $E_p$, divided by $t^\prime_{dyn}$,  are plotted for first-order acceleration (F1), from eq.\ (14), and second-order acceleration (F2), from eq.\ (15), by solid and dotted lines, respectively. The progressively lighter lines take $f_1,f_2 = 1, 10, 10^2, 10^3,$ and $10^4$, respectively. The assumed parameters permit acceleration of protons to $\approx 8\times 10^{19}$ eV with maximum efficiency. Protons with energies greater than this energy, shown by the cross-hatched region, are not allowed by the Hillas condition.

The maximum energy of escaping particles is limited by the escape timescales. Fig.\ 6 also shows the ratio of the diffusive escape timescale to the dynamical timescale due to gyroresonant pitch-angle scattering with Alfv\'enic turbulence, using the expression
\begin{equation}
{t^\prime_{\rm F2,esc}\over t^\prime_{dyn}} \cong \max[1,{\pi\over 8}(q-1)(2-q)(4-q)\zeta \left( {E\over QB^\prime c\Gamma^2 t_{var}}\right)^{q-2}]\;
\end{equation}
\cite{dml96}.  This ratio cannot be less than unity because particles cannot escape on timescales shorter than $t_{dyn}$. We set the coefficient ${\pi\over 8}(q-1)(2-q)(4-q)\zeta$ equal to unity in order to give the longest possible escape timescale through gyroresonant diffusion in the dashed line denoted ``F2,esc" in Fig.\ 6. { (A second dashed line assumes a factor of 10 more rapid escape, corresponding to an order-of-magnitude reduction in the plasma turbulence energy density.)} But note that Eq.\ (16) assumes a picture where there are open magnetic field lines along which the particles diffuse and escape from the jet plasma. A more realistic picture for blazars might be Bohm diffusion in a randomly oriented magnetic field. The ratio of the Bohm diffusion timescale to the dynamical timescale takes the simple form 
\begin{equation}
{t^\prime_{\rm Bohm,esc}\over t^\prime_{dyn}} \cong \max[1,\left( {E\over QB^\prime c\Gamma^2 t_{var}}\right)^{-1}]\;,
\end{equation}
and is shown in Fig.\ 6 by the dot-dashed line denoted ``Bohm,esc." 

Maximum particle energy is also limited by radiative losses. Fig.\ 7 shows the timescale for photohadronic energy losses with scattered 
accretion disk, BLR, and IR torus photons, for parameters of a flaring FSRQ.
In this case, photopion losses will limit proton acceleration to $\lesssim 10^{17}$ eV in first-order acceleration if $f_1 \cong 10^5$ (i.e., an acceleration efficiency of 0.001\%), assuming Bohm diffusion.  In comparison, a value of $f_2 \gtrsim 1000$  for second-order acceleration limits proton acceleration to $\approx 10^{16}$ eV, assuming Bohm diffusion.
The acceleration efficiency is difficult to estimate in either case, and depends on the uncertain level of turbulence and the Alfv\'en speed. But it is important to note that in both first- and second-order Fermi acceleration, a characteristic maximum proton energy of $\approx 10^{16}$ eV, i.e., $\gamma_{pk} \sim 10^7$, is a consequence of energy losses off the BLR radiation when the acceleration efficiency is sufficiently small and photohadronic losses with the BLR are sufficiently large. This feature of particle acceleration could explain the apparent cutoff in multi-PeV neutrino events observed with IceCube.

Second-order Fermi acceleration gives a curving accelerated particle distribution resulting from diffusive acceleration, but inefficient first-order acceleration will also produce a curving spectrum with a spectral cutoff due to photohadronic losses with BLR photons. As noted previously, we chose a value of the log-parabola parameter $b\cong 1$  based on fits to nonthermal electron synchrotron spectra in blazars \cite{der14}. In principle, $b$ can be derived by comparing the proton distribution formed as a consequence of particle acceleration, energy-loss and escape, or by directly using theoretical particle spectra resulting from scenarios involving Fermi acceleration \cite{bld06,sp08,matm12,der13,asa14}. 

\subsection{Diffuse neutrino intensity}

The large directional uncertainty makes association of IceCube neutrinos with point sources difficult, particularly for shower events. Although a truly diffuse cosmogenic origin of the IceCube neutrinos is ruled out, as noted in the Introduction, the PeV neutrinos may be associated with large fluence FSRQs.
No convincing associations have been made (see Section 6.1), and we can speak of the PeV neutrinos as ``diffuse," even though they may be due to the superposition of blazars that are not individually resolved by IceCube but resolved by Fermi.

PeV neutrinos originating from FSRQs { will unavoidably be accompanied by $\gamma$ rays}, 
so that  $\gamma$-ray fluence provides perhaps the best index to search for high-energy neutrino
sources.  FSRQs make a significant contribution to the $\gamma$-ray background \cite{it09,aje12}.  
A calculation of the ``diffuse" neutrino intensity based on the blazar sequence is presented in Ref.\ \cite{mid14}. 
Besides searches for neutrinos from known point-source blazar directions, another method to test this model 
is to compare the probability for high $\gamma$-ray fluence FSRQs to be found in PeV neutrino directional 
error ellipses with sources described by the same fluence distribution that are distributed 
randomly on the sky. With only three $> 1$ PeV neutrinos so far reported with IceCube, 
such tests are as yet statistically challenging \citep[e.g.,][]{kra14}, but will become more promising as exposure, 
both with IceCube and Fermi,  grows with time.

Regarding all 28 \cite{IceCube2013} and now 37 \cite{aar14} excess neutrino events, the implied intensity of the excess IceCube neutrino flux is at the level of $\sim 3\times 10^{-8}$ GeV/cm$^2$-s-sr \cite{IceCube2013,wax13}. The PeV neutrinos alone contribute more than 50\% of this intensity. This can be compared with the integrated $\gamma$-ray intensity of FSRQs measured with Fermi-LAT between 100 MeV and 100 GeV \cite{ack11}. The cumulative energy-flux ($\Phi$) distribution measured in the range of $10^{-11} \leq \Phi$ (erg/cm$^{2}$-s) $\leq 10^{-9}$ implies  an FSRQ $\gamma$-ray intensity $\approx 5\times 10^{-7}$ GeV/cm$^2$-s-sr, which is a lower limit given that FSRQs with $\Phi$ outside this range are not counted. Because neutrino production will unavoidably produce $\gamma$ rays with comparable intensity that, though generated at very high energy, can cascade into the LAT energy range, this estimate indicates that $\approx 10$\% of the FSRQ $\gamma$-ray emission could be produced by hadronic jet processes. { Diffuse 100 TeV -- PeV neutrinos from one-zone models of BL Lac objects require low Doppler factors that would imprint strong $\gamma\gamma$ opacity features on the SED, unlike the observed SEDs of BL Lac objects.}

\section{Conclusions} 

In order to avoid overproduction of $\gg$ PeV neutrinos by cosmic-ray protons in FSRQs, a typical FSRQ proton spectrum (reflecting an average over many sources) that softens at energies $\gtrsim 100$ PeV is required.  The proton distribution could be in the form of a broken power law or an exponentially cutoff power law, but here we consider a log-parabola function. The lack of high-energy neutrinos can then be explained if $b\cong 1$ and the principal Lorentz factor $\gamma_{pk} \lesssim 10^{8}$.  The corresponding proton energies, $\lesssim 10^{17}$ eV, are well below energies needed to explain the UHECRs. 

Indeed, FSRQs and their off-axis counterparts cannot be the principal sources of UHECRs extending to $\approx 10^{20}$ eV, as they are not found within the GZK radius, and their $\gamma$-ray energy production rate per unit volume, if comparable to the required UHECR emissivity, is inadequate in the local universe (unlike the case of BL Lac objects). The presence of strong external radiation fields in the inner jets of FSRQs may inhibit acceleration of protons and ions to ultra-high energies, in the same way that the mean electron Lorentz factors in FSRQs are much less than those in BL Lac objects \cite{fos98,ghi98} due, apparently, to radiative cooling. This behavior of the electron distribution, which helps explain the blazar sequence relating the peak synchrotron luminosity with the frequency of the synchrotron peak, would have analogous behavior for hadrons, in accord with the hypothesis that BL Lac objects are the sources of the UHECRs. Intermediate and high-synchrotron peaked BL Lac objects are then predicted to be candidates for detection of EeV neutrino point sources by the Askaryan Radio Array  \cite{ara12}. A separate study is required to demonstrate whether a superposition of curved UHECR injection spectra from blazars at various redshifts reproduces the UHECR spectrum. 

In related work \cite{mid14}, we considered the latest blazar gamma-ray luminosity function in order to derive the diffuse intensity of neutrinos made in blazar jets, which is dominated by production in FSRQs due to their strong external radiation fields. There we find a similar result \cite{mid14} using a power-law cosmic-ray spectrum to explain UHECRs.

Our calculations of neutrino spectra from FSRQs show a hardening below $\approx 1$ PeV from the spectrum of decaying pions. Superposition of emission from blazars at various redshifts is not sufficient to conceal this low-energy cutoff \cite{mid14}. Evidence for a suppression of neutrinos below $\approx 1$ PeV would support this model, but the existence of a gap in the neutrino spectrum at these energies is not statistically significant \cite{IceCube2013,aar14}. As IceCube exposure grows, our model will be tested by measuring the $\gtrsim 100$ TeV neutrino spectrum. If FSRQs are the sources of the IceCube PeV neutrinos, a second component is unavoidably required to explain the  $\lesssim 300$ TeV neutrino events, for which star-forming galaxies, galaxy clusters and groups, or a
higher prompt atmospheric neutrino background provide plausible
explanations. Lack of evidence for a gap between a few hundred TeV and $\approx 1$ PeV in the IceCube neutrino spectrum would instead provide evidence for a single-source model of neutrino production, for example, a nuclear production model 
where neutrinos originate from cosmic-ray reservoirs (such as galaxies and galaxy assemblies) with typical cosmic-ray spectra described by a power-law with index harder than $\sim 2.2$ and a break or cutoff near 100 PeV  \cite{mur13,tam14}. By comparison,
in the model studied here where PeV neutrinos are produced by photopion processes in the inner jets of FSRQs,  a suppression of the neutrino flux below $\approx 1$ PeV is predicted, and can furthermore be tested by identifying  high $\gamma$-ray fluence FSRQs in PeV neutrino error boxes. 

\vskip 0.2in
\noindent{\bf Acknowledgments}
We wish to thank J.\ Becerra, E.\ Blaufuss, J.\ Finke, A.\ Kusenko, B.\ Lacki, B.\ Lott, A.\ Reimer, 
and K.\ Schatto  for discussions and correspondence.
{ We would like to acknowledge the very useful report of the referee, which helped clarify the issues surrounding this model.
} The work of C.D.D.\ is supported by the Office of Naval Research. K.M.\ is supported by NASA through Hubble Fellowship, Grant No. 51310.01 awarded by the Space Telescope Science Institute, which is operated by the Association of Universities for Research in Astronomy, Inc., for NASA, under Contract No. NAS 5-26555. 

\appendix

\section{Synchrotron Self-Absorption (SSA) Frequency in Blazars}

{ 
Synchrotron spectra of blazars may be self-absorbed at low radio frequencies.
Following Ref.\  \cite{der14}, the SSA optical depth of magnetoactive plasma with average
magnetic field $B^\prime$ and an electron Lorentz-factor $\gamma_e^\prime$ distribution described by the log-parabola function $\gamma_e^{\prime 2}N_e^\prime(\gamma_e^\prime) 
= K y^{-b \log y}$, where $y \equiv \gamma_e^\prime/\gamma^\prime_{pk}$, is given 
in the $\delta$-function approximation for photons with comoving dimensionless energy $\epsilon^\prime$  by
\begin{equation}
\tau_{\ep} = {8\pi\over 9} \,{r_b^\prime \lambda_{\rm C} r_e\over m_ec^2 I_1(b)} \,{u^\prime_{B^\prime} u^\prime_e
\over u_{cr}}\,{{\cal F}(\hat y)\over \epsilon^{\prime 3}}\;.
\label{tauepsilon}
\end{equation}
Here $u_e^\prime$ is the nonthermal electron energy density, $u^\prime_{B^\prime} = B^{\prime 2}/8\pi$, $u_{cr} = B_{cr}^2/8\pi$ is the critical field energy density,
$B_{cr} = 4.41\times 10^{13}$ G, $r_b^\prime \cong c \delta_{\rm D} t_{var}$, $I_1(b) = \sqrt{\pi\ln 10/b}$, $\lambda_{\rm C} = h/m_ec$ is the 
Compton wavelength, 
and 
\begin{equation}
{\cal F}(\hat y) \equiv (1 + {b\over 2} \log \hat y )\, \hat y^{-b \log \hat y}\;,
\label{Fhaty}
\end{equation}
with  
\begin{equation}
\hat y \equiv {\sqrt{ { \epsilon_{SSA}/ 2 \varepsilon_{B^\prime} \delta_{\rm D} } }\over \gamma_{pk}^\prime }\;.
\label{haty}
\end{equation} 
Note that ${\cal F}(\hat y = 1) = 1$.

Defining the SSA frequency by $\tau_{\ep_{SSA}} = 1$ gives 
\begin{equation}
 \epsilon_{SSA} = \delta_{\rm D} \ep_{SSA} = \delta_{\rm D}^{4/3} \left[ {8\pi \lambda_{\rm C} r_e c t_{var}\over 
9 m_e c^2 I_1(b)}\,\left({\zeta_e u^{\prime 2}_{B^\prime} \over u_{cr}}\right) {\cal F}(\hat y)\right]^{1/3}\;.\;
\label{epsilonSSA}
\end{equation}
Relating $u_e^\prime= \zeta_e u^\prime_{B^\prime}$ through the parameter $\zeta_e$, one can show that the equipartition condition $\zeta_e= 1$ is close to the minimum jet power condition \citep{der14}.
Note that eq.\ (\ref{epsilonSSA}) is transcendental through the dependence of $\hat y$ on $\epsilon_{SSA}$.

Eq.\ (\ref{epsilonSSA}) gives the SSA frequency 
\begin{equation}
\nu_{SSA} \cong 140 {\rm ~GHz} \;\left( {\zeta_e t_3\over I_1(b)}\right)^{1/3}\;B^{\prime 4/3}({\rm G})\,\left({\delta_{\rm D}\over 10}\right)^{4/3}\,,\;
\label{nuSSA}
\end{equation}
dropping the slowly varying factor ${\cal F}^{1/3}(\hat y)$ and defining $t_n = t_{var}({\rm s})/10^n$ s.
It can easily be seen that for BL Lac objects and GRBs, which 
have equipartition magnetic fields $B^\prime_{eq} \sim 100$ mG and $\sim 10$ G \citep{der14}, respectively,
$\epsilon_{SSA} \ll \epsilon_{pk}$. FSRQs have $B^\prime_{eq} \sim$ few G, and with $\epsilon_{pk} \sim 10^{-7}$,
have $\epsilon_{pk} \sim \epsilon_{SSA}$. SSA effects are never important for PeV
neutrino production, though could be important for EeV neutrino and $\gamma$-ray production in FSRQs. 
The effects of SSA hardly change the calculation of the SSC component.
}

\section{Photopion Production Efficiency and Neutrino Luminosity}

{ 

We optimize neutrino luminosity for neutrinos with energy $E_\nu = \chi m_p \delta_{\rm D}\gamma^\prime_p$
formed from protons with Lorentz factors $\gamma_p =\delta_{\rm D}\gamma^\prime_p = E_\nu/\chi m_p$, $\chi \cong 0.05$
\citep[see also][]{drl07}. Assume that a fraction $k_p$ of the
jet power $L_{j,*}$ in the galaxy/black-hole frame is transformed into nonthermal proton power $L_{p,*} = k_p L_{j,*}$.
Because $dt^\prime = dt_*/\Gamma$, ${\cal E}_p^\prime = {\cal E}_{p,*}/\Gamma$, $L_p^\prime = L_{p,*}\cong L_p/\Gamma^2$
(for a blast wave).
Assuming that the target photons are isotropically distributed in the fluid frame, $L_{\nu} = \delta^{4} L^\prime_\nu$ (for a blob).
Further, $L_\nu^\prime = \chi L_p^\prime \min(1,\eta_{\phi\pi})$, where $\eta_{\phi\pi} = \eta_s I_s(\bar x)$,
and $\eta_s \equiv K_s/\delta_{\rm D}^4$ is 
given by eq.\ (\ref{etas}), where $\bar x = \delta_{\rm D}\sqrt{ {\bar \epsilon_{thr}\chi m_p/ (2 E_\nu \epsilon_{pk})} }$. 

From inspection, optimal neutrino production from a black-hole jet source occurs for $\bar x \cong 1$, so 
at $\bar x \cong 1$, $I_s(\bar x) \cong 1$, and the optimal Doppler factor in terms of neutrino 
production is defined by 
$$\hat \delta_{\rm D} \cong {11 L_{48}^{1/4}\over t^{1/4}_{var}({\rm s})f_0^{1/4}10^{1/16b}\epsilon_{pk}^{1/4} } \;\arrowsim\; $$
\begin{equation}
9.6 \left({L_{48}\over f_0 t_{var}({\rm s}) \epsilon_{pk}}\right)^{1/4}\;\arrsim\;   12.6 \left({L_{48}\over t_{var}({\rm s}) \epsilon_{pk}}\right)^{1/4}.\;
\label{hatDoppler}
\end{equation}

For a blast-wave geometry, $f_0 = 1$, and $f_0 = 1/3$ for a blob \citep{der14}. 
Defining $L_n = L/10^n{\rm~erg~s}^{-1}$ and $\epsilon_{n} = \epsilon_{pk}/10^n$,
\begin{equation}
\hat \delta_{\rm GRB} \cong 170 \left({ L_{52}\over t_{var}(0.1{\rm~s})\epsilon_{pk} }\right)^{1/4}\;,\;
\label{hatdeltaD}
\end{equation}
and the condition $\bar x \cong 1$ implies that the efficiency is maximized for neutrinos formed
at energy $E_\nu \cong \delta_{\rm D}^2\bar\epsilon_{thr} \chi m_p/2\epsilon_{pk}$ which, for 
GRBs, implies 
\begin{equation}
E_{\nu,{\rm GRB}} \cong 270 \sqrt{L_{52}/t_{var}({\rm 0.1~s})\epsilon_{pk}^3} {\rm ~TeV}\;.
\label{EnuGRB}
\end{equation}
For an HSP BL Lac object, the most luminous neutrino fluxes from internal processes are found when
\begin{equation}
\hat \delta_{\rm BL} \cong 4.0\,\large( {L_{46}\over t_3 \epsilon_{-3}}\large)^{1/4}\,,
\; E_{\nu,{\rm BL}} \cong 150 \sqrt{L_{46}\over t_3\epsilon_{-3}^3}\;{\rm TeV}\;.\;
\label{hatdeltaDBL}
\end{equation} 
With such low Doppler factors, GeV -- TeV $\gamma$ rays would be strongly attenuated by $\gamma\gamma$ 
pair production \citep{drl07}, whereas BL Lac objects show no indication of internal $\gamma\gamma$ absorption.
Furthermore, such low Doppler factors correspond to systems far from equipartition \citep{der14},
which are disfavored energetically.

For the low synchrotron-peaked FSRQs, with $\e_{pk}\cong 10^{13}{\rm Hz}$, 
or $\epsilon_{pk} = 10^{-7}\epsilon_{-7}$,
\begin{equation}
\hat \delta_{\rm FS} \cong 71\,\large( {L_{48}\over t_4 \epsilon_{-7}}\large)^{1/4}\,,
\; E_{\nu,{\rm FS}} \cong 4.7\times 10^{20} \sqrt{L_{48}\over t_4\epsilon_{-7}^3}\;{\rm eV}.\;
\label{hatdeltaDFSRQ}
\end{equation}
The class of intermediate synchrotron-peaked blazars with $10^{-6}\lesssim \epsilon_{pk} \lesssim 10^{-5}$
would be favored to make EeV neutrinos by this logic. Provided that a broadband spectrum 
of protons is accelerated with a number index of $\approx 2$, specific values of Doppler factor
for GRBs and HSP BL Lac objects optimize $\sim 100$ TeV neutrino production.
}

\section{Threshold Lorentz Factor for Photopion Production from Direct Accretion-Disk Radiation}

Consider accretion-disk photons passing through a plasma jet moving outward with bulk Lorentz factor $\Gamma = 1/\sqrt{1-\beta^2}$ along the axis of the accretion disk. In the jet fluid frame, the threshold condition for photopion production by ultra-relativstic protons is simply $\gamma_p^\prime\epsilon^\prime(1-\mu^\prime)> \bar\epsilon_{thr}$, using notation of the Section 2. Here $\epsilon^\prime = \Gamma\epsilon(1-\beta\mu)$, and $\mu^\prime = (\mu-\beta)/(1-\beta\mu)$. The term $\mu = r/\sqrt{r^2 + R^2}$ is the cosine angle of the photon emitted by the accretion disk at radius $R$ from the nucleus that intercepts the jet at distance $r$ along the polar axis of the accretion disk. 

The mean photon energy radiated from an optically thick Shakura-Sunyaev accretion disk is $m_ec^2 \epsilon(\tilde R) \cong 77 q \tilde R^{-3/4}$ eV, where $q = (\ell_{\rm Edd}/M_9 \eta)^{1/4}$, $10^9 M_9 M_\odot$ is the black hole mass, $\ell_{\rm Edd}$ is the ratio of the radiant luminosity to the Eddington luminosity, and $\eta\cong 0.1$ is the efficiency of the accretion disk for converting accretion power into luminosity  \cite{ds02}.The tildes refer to quantities measured in units of the gravitational radius $r_g = GM/c^2$. Writing $\epsilon(\tilde R) \cong 1.5\times 10^{-4} q \tilde R^{-3/4}$ gives the threshold condition 
\begin{equation}
\gamma_p^{thr} \cong \Gamma\gamma_p^\prime = {A x^{3/4}\over 1-{\beta/\sqrt{1+x^2}}}\;, 
\end{equation} 
for photopion production by a proton  with Lorentz factor $\gamma_p^{thr}$. Here $A\equiv {\bar \epsilon_{thr}/ 1.5\times 10^{-4}q}$.  Differentiating gives the minimum value of $\gamma_p$ which, for $\tilde R\gg 1$, is at $\gamma_p^{thr} \cong 8\times 10^6 \tilde r^{3/4}/q$. This can be rewritten to give the minimum energy $E_p^{thr} \cong 2.4\times 10^{17}q^{-1} (r/100r_g)^{3/4}$ eV, which is independent of $\Gamma\gg 1$.  A value of $q\cong 0.1$ gives typical temperatures of the optically-thick accretion disk in FSRQs, which leads to even higher values of $E_p^{thr}$. Unless we consider extreme inner jet models with $r\ll 100 r_g$, we can therefore neglect photopion production from the direct accretion-disk radiation field for the production of PeV neutrinos.


\begin{thebibliography}{}
\bibitem{footnote}The derivation 
depends on whether the proton is assumed to escape from the jet 
 with $\gamma_p\cong \Gamma\gamma_p^\prime$ or $\gamma_p \cong 2\Gamma\gamma_p^\prime$; 
here we assume the former relation.
\end{thebibliography}






\end{document}